\documentstyle[preprint,aps]{revtex}

\newcommand{\vecf}{\!\!\mbox{ \boldmath $F$}}
\newcommand{\veca}{\!\!\mbox{ \boldmath $A$}}
\newcommand{\vect}{\!\!\mbox{ \boldmath $\tau$}}
\newcommand{\vecsigma}{\!\!\mbox{ \boldmath $\sigma$}}

\begin{document}

\draft

%
%

\title{Dilatonic Black Holes with Gauss-Bonnet Term}
\author{Takashi Torii\thanks{electronic 
mail:torii@cfi.waseda.ac.jp}, 
Hiroki Yajima\thanks{electronic mail:695L1079@cfi.waseda.ac.jp}
and
Kei-ichi Maeda\thanks{electronic mail:maeda@cfi.waseda.ac.jp}}
\address{Department of Physics, Waseda University,
Shinjuku-ku, Tokyo 169, Japan}
\date{\today}
\maketitle
\begin{abstract}
\baselineskip=12pt

We discuss black holes in an effective theory derived from a 
superstring 
model, which includes a dilaton field, a gauge field and the 
Gauss-Bonnet  term. Assuming U(1) or SU(2) symmetry for the gauge 
field, we find four types of spherically symmetric solutions, i.e.,
a  neutral, an electrically charged, a magnetically charged and a 
``colored'' black hole, and discuss their thermodynamical properties
and fate via the Hawking evaporation process. For  neutral and 
electrically charged  black holes, we find critical point and a
singular end point. Below the mass corresponding to the critical 
point,  no
solution  exists, while the curvature on the horizon  diverges and a
naked singularity appears at the singular point. A cusp structure
in the mass-entropy diagram is found at the critical point and
black holes on the branch between the critical and
singular points become unstable.  For magnetically charged and 
``colored" black holes, the solution becomes singular just at the
 end point with a finite mass. Because the black hole
temperature is always finite even at the critical  point or the singular
point,  we may conclude that 
the  evaporation process will not be stopped even at the critical
point or the singular point, and
the black hole will move to a dynamical
evaporation phase or a  naked singularity will appear. 

\end{abstract}

%
%
\section{Introduction}
\label{sec:Introduction}

One of the most fascinating dreams for all physicists is the 
unification of all fundamental forces, i.e., electromagnetic, weak, 
strong and gravitational interactions. The electromagnetic and 
weak interactions are successfully unified in the Weinberg-Salam 
theory. The strong interaction is described by quantum 
chromodynamics (QCD) and is likely to be unified with the 
Weinberg-Salam theory into a grand unified theory in the context of 
gauge 
theory. The gravitational interaction, however, is not yet included, 
in spite of a great deal of effort. The most promising candidate for 
a unified theory of all interactions is a superstring theory, which 
may unify everything without any divergences.

Although such a unified theory may become important in the strong
gravity regime, however, little work has been done on 
extreme situations such as a black hole or the early universe. 
Since methods to study the strong gravity regime in a string theory
are not
well developed, most analysis have been performed by
using an effective field theory inspired by  a string theory,
which contains the leading or next to leading  order terms
of the inverse string tension
$\alpha'$. One such application  is string cosmology. Some puzzles
in Einstein  cosmology might be solved with a string theory. For
example, while the  singularity theorem demand that the universe
has an initial singularity in the Einstein gravity, a string inspired 
model
can remove it and provide a non-singular cosmology\cite{ST,EM}.

Another application is black hole physics. The first study was 
made by Gibbons and one of the present authors in the 
Einstein-Maxwell-Dilaton (EMD) system\cite{GM} and the same solution 
was also
discussed in a different coordinate system in ref.\cite{Garfinkle}. They 
found a static spherically symmetric black hole solution (GM-GHS 
solution) with  dilaton hair. Since the dilaton hair cannot appear 
without  electromagnetic hair, it is classified as secondary 
hair. After this work, many solutions were discussed in various 
models. Thus the following question naturally arises:  How are the
black  hole solutions affected by the next to leading order term in
$\alpha'$, in particular by 
 the higher-curvature term?  This was first considered 
independently of a string theory. Wheeler analyzed the effect of the
Gauss-Bonnet (GB) term without a dilaton  field\cite{Wheeler}. When
the dilaton field is absent, the GB term  does not give any
contribution  in a four dimensional spacetime
because  it becomes a surface term and gives a topological
invariant. 
Then  he studied
 black holes  in spacetime with more than four dimensions.  
These solutions are called dimensionally continuum black 
holes\cite{dimcon}. Callan et al.\cite{CMP} discussed black hole 
solutions in the theory with a higher-curvature term $R_
{\mu\nu\rho\sigma} R^{\mu\nu\rho\sigma}$ and a dilaton field,  and
Mignemi and Stewart\cite{MS} took both the GB term and a dilaton
field into account in four dimensional spacetime. In their  work,
field variables are expanded by the inverse string tension 
$\alpha'$ and the first order terms of $\alpha'$ are taken into 
account. Using this perturbation, they constructed  analytic 
solutions.  There are also some studies which clarify the effect of an
axion field as well as  the dilaton field, a U(1) 
gauge field and the GB term\cite{AGB1,AGB2,AGB3}. 
Either dyon solutions or axisymmetric  stationary 
solutions are analyzed 
because  the axion field becomes trivial if the gauge field does
not have both electric and magnetic charges and the
spacetime is static and spherically symmetric. 
In all these models containing a higher-curvature term, 
a perturbative approach was used. Assuming the  GB
higher curvature term, recently, Kanti et. al. 
calculated a
neutral solution without such a perturbation and found 
interesting  properties\cite{KMRTW}. More recently Alexeyev and 
Pomazanov also
discussed the internal structure of these solutions\cite{AP}.

When we turn to another aspect of a unified theory, we find a
non-Abelian gauge field.  One of the most important facts about black
holes with a
non-Abelian gauge field is that we have the so-called colored black
hole, which  was found in Einstein-Yang-Mills 
system\cite{Vol} soon after the discovery of a particle-like 
Bartnik-McKinnon(BM) solution\cite{Bar}. These solutions can exist by a
balance between the attractive  force of gravity and the repulsive
force of the Yang-Mills (YM) field.  If gravity is absent, such
non-trivial structures cannot exist. In this sense,  these solutions
  are of a
new type. Although both the BM particle  and the colored black hole are
found to be unstable against radial  linear perturbation\cite{Gal},
they showed us a new aspect of black  hole physics and forced us to
reconsider the black hole no-hair  conjecture.

After these solutions were discussed, a variety of self-gravitating 
structures and 
black hole solutions with a non-Abelian field were found in static 
spherically symmetric spacetime\cite{Tor2,Tor3,Tach}.  The 
Skyrmion\cite{Sky,Dro} or the Skyrme black hole\cite{Luc,Dro,Tor} in 
Einstein-Skyrme system, the particle solution with a massive 
Proca field or the Proca black hole\cite{Gre} in Einstein-Proca 
system, the monopole\cite{'tH,LNW,Bre,Tach} or the black hole in a 
monopole\cite{LNW,Bre,Aic,Tach} in the Einstein-Yang-Mills-Higgs
(real triplet) system and the sphaleron\cite{Das,Gre} or sphaleron 
black hole\cite{Gre} in Einstein-Yang-Mills-Higgs(complex doublet) 
system have been discovered. A particular class of the Skyrme black 
hole, the 
Proca black hole and the monopole black hole turns out to be stable 
against radial perturbations. In particular the monopole black 
hole can be  a counterexample of the black hole no-hair 
conjecture\cite{Tach}, because it is highly stable and is formed
through the Hawking evaporation process from the Reissner-Nordstr\"
{o}m
black hole. We also investigated the dilatonic BM  particle and the
dilatonic colored black hole solution in the 
Einstein-Yang-Mills-Dilaton (EYMD) system\cite{Tor,dil}, which are 
direct extensions of the GM-GHS solution. 

Although those non-Abelian fields may be expected in some unified
theories, such black holes must be very small and then some other
contributions such as the GB term  and/or the moduli field may also
play an important role in their structure, if  the fundamental theory
is described by a string model.  Donets and  Gal'tsov showed that a
particle-like solution does not exist in the   EYMD system with the
GB term\cite{DG}.  Then, they assume a  numerical constant
$\beta$ in front of the GB term, where
$\beta =1$  corresponds to the effective sting theory. They showed
that there  is a critical value $\beta_{cr} =0.37$, beyond which no
particle-like  solution exists. However, we expect that a black
hole solution  can exist in this system, so we also study the
case of a SU(2) YM field. 

In this paper, then, we study black holes in a theory inspired by a
string theory, i.e., in a model with a dilaton field, a gauge field,
and the GB curvature term, and discuss their properties.  This paper is
organized as follows. We outline the a model and field  equations 
in section
2, and present various types of new solutions  (neutral, electrically
charged, magnetically charged and ``colored''  black holes) in section
3. In section 4 we study the  thermodynamical properties of those black
holes. Section 5  includes discussions and some remarks.

%
%
\section{A Model and Field Equations}
\label{sec:Model-Equations}

We shall consider  
the model given by the action
\begin{equation}
  S = \int d^4x \sqrt{-g}\left[\frac1{2\kappa^2}R
        -\frac1{2\kappa^2}\left( \nabla \phi\right)^2
        -\frac16e^{-2\gamma\phi} H^2
        +\frac{\alpha '}{16\kappa^2} \mbox{e}^{-\gamma \phi}
        \left(\hat{R}^2 - {\rm Tr}\vecf^2  \right)
        \right],
\label{2-10} 
\end{equation}
where $\kappa^2=8\pi G$, and discuss a spherically symmetric,
static solution. This type of action comes from low-energy limit of
the heterotic string theory\cite{Gross}.
The dilaton field is $\phi$ and $\gamma =\sqrt{2}$ is the coupling
constant of the dilaton field to the gauge field. $H$  is a three form
expressed as
\begin{equation}
  H = dB + \frac{\alpha'}{8\kappa} \left(\Omega_{3L}-\Omega_
{3Y}\right),
\label{2-15} 
\end{equation}
where $B_{\mu\nu}$ is the antisymmetric field in the 
gravitational multiplet. $\Omega_{3L}$ and $\Omega_{3Y}$ are the 
Lorentz and gauge Chern-Simon terms respectively;
\begin{eqnarray}
  \Omega_{3L} & = & {\rm tr} \left( \omega \wedge R 
              - \frac13 \omega \wedge \omega \wedge \omega \right), 
\\
  \Omega_{3Y} & = & {\rm Tr} \left( \veca \wedge \vecf 
              - \frac13 \veca \wedge \veca \wedge \veca \right),
\label{2-17} 
\end{eqnarray}
where the traces are taken over the Lorentz and gauge indices. 
$\omega_{a\mu\nu} =  e_{(a)}^\rho e_{\mu}^{(b)} \nabla_\rho e_{\nu
(b)}$ is the spin  connection with vierbein $ e_{\mu}^{(a)}$.
Using the dual of the Bianchi identity, we can rewrite a part of the
action
\begin{eqnarray}
        -\frac16 e^{-2\gamma\phi} H_{\mu\nu\rho} H^{\mu\nu\rho} & = &
           -\frac{1}{4\kappa^2} e^{2\gamma\phi} 
           \partial_{\mu} b_{KR} \partial^{\mu} b_{KR}
          \nonumber \\ && ~~~
 +\frac{\alpha'}{32\kappa^2} \left[ b_{KR}
\epsilon^ {\rho\sigma\mu\nu}
             \left(R_{\alpha\beta\rho\sigma} R_{\mu\nu}^
{\;\;\alpha\beta}
             +{\rm Tr} \vecf_{\rho\sigma}\vecf_{\mu\nu} \right) 
\right].
\label{2-57}
\end{eqnarray}
 The pseudoscalar  field $b_{KR}$ is the
 Kalb-Ramond axion.
If a spacetime is static and spherically symmetric and the gauge 
field does not have dyons but only an electric or a magnetic charge,
which is the situation we will discuss here, then the second term of 
equation (\ref{2-57}) vanishes. Then the axion field can be 
regarded as a massless scalar field. 
Such a scalar field, however,  
is trivial because of the black hole no hair theorem\cite{Bek}.
Hence we can drop the axion field here.

$\vecf$ is the field strength of the gauge field expressed by its 
potential $\veca$. If it is a $U(1)$ gauge field (the
electromagnetic field), then we consider only a electrically
or magnetically charged black hole, i.e., 
\begin{equation}
\veca = a dt ~~~{\rm and} ~~~~\vecf = {da \over dr} ~ dr\wedge dt .
\end{equation}
For $SU(2)$ gauge field, we assume the Witten
ansatz\cite{Wit}, which  is the most generic form of a spherically
symmetric SU(2) YM  potential. The YM potential becomes
\begin{equation}
       \veca =a\vect_r dt+b\vect_r dr
           +\left[ d\vect_{\theta} -(1+w) \vect_{\phi} \right] 
           d\theta
           +\left[ (1+w)\vect_{\theta} +d\vect_{\phi} \right]
           \sin \theta d \phi ,
\label{3-30}
\end{equation}
where $a$, $b$, $d$ and $w$ are functions of  time and the radial
coordinates, $t$ and  $r$. We have adopted the polar coordinate 
description
$(\vect_r, \vect_{\theta}, \vect_{\phi})$, i.e.
\begin{eqnarray}
         \vect_r & = & \frac1{2i} [\vecsigma_1 \sin \theta \cos \phi 
         + \vecsigma_2 \sin \theta \sin \phi 
         + \vecsigma_3 \cos \theta] ,
\label{3-40}  
\\
         \vect_{\theta} & = & \frac1{2i} [\vecsigma_1 \cos \theta 
\cos \phi 
         + \vecsigma_2 \cos \theta \sin \phi 
         - \vecsigma_3  \sin \theta] ,
\label{3-50}  
\\
         \vect_{\phi} & = & \frac1{2i} [-\vecsigma_1 \sin \phi 
         + \vecsigma_2 \cos \phi] ,
\label{3-60}  
\end{eqnarray}
whose commutation relations are
\begin{equation}
     \left[ \vect_a, \vect_b \right] =\vect_c 
  \hspace{10mm} a, b, c, = r, \theta , ~{\rm or }~~\phi. 
\label{3-70}
\end{equation}
Here $\vecsigma_i \hspace{3mm} (i=1,2,3)$ denote the Pauli spin 
matrices.
We can  eliminate  $b$ using a residual gauge freedom. 
In the static case,  the part of the YM equations is integrated as $d=
Cw$ where $C$ is an integration constant. We can set $C=0$ i.e. $d 
\equiv 0$ without loss of generality. The remaining functions $a$ 
and $w$ depend only on the  radial coordinate $r$. As a result, we 
obtain a simplified spherically symmetric YM 
potential as  
\begin{equation}
        \veca = a(r) \vect_r dt -\left[1+w(r)\right]\vect_{\phi} 
d\theta
       +\left[1+w(r)\right]\vect_{\theta} \sin \theta d\phi.
\label{3-80}
\end{equation}
Substituting this  into $\vecf =d\veca +\veca \wedge \veca$,
we find the field strength:
 \begin{eqnarray}
          \vecf  &=& {d a \over d r} \vect_rdr\wedge dt
                 +{d a \over d r} \vect_{\phi} dr\wedge d\theta
                 +{d a \over d r} \vect_{\theta} dr\wedge \sin
\theta d\phi      
               \nonumber\\
       & & \mbox{ } -\left(1-w^2\right) \vect_rd\theta \wedge \sin 
\theta d\phi
                 +aw\vect_{\theta} dt \wedge d\theta
                 +aw\vect_{\phi}dt\wedge \sin \theta d\phi.
\label{3-90}
\end{eqnarray}
Comparing (\ref{3-90}) with the field strength of the U(1) 
gauge field, we find that $a$ and $w$ play the roles of an electric 
and a magnetic potentials, respectively. This expression can be
used for U(1) gauge field if we formally set 
$\vect_r =1$ and $\vect_{\theta}=\vect_{\phi}=0$.

The GB term, $\hat{R}^2$, is defined by 
\begin{equation}
  \hat{R}^2 = R_{\mu \nu \rho \sigma} R^{\mu \nu \rho \sigma}
       - 4R_{\mu \nu} R^{\mu \nu} + R^2.
\label{2-20} 
\end{equation}
This combination is introduced to cancel anomalies and has the 
advantage that the higher derivatives of metric functions do not 
appear in the field equations. 
Setting $\alpha'/\kappa^2 = 1/\pi g^2$, 
$g$ is regarded as  a gauge coupling constant. A numerical 
constant  $\beta$ is introduced in Ref. \cite{DG} to find a
non-trivial particle-like solution, but we fix it to be unity
because we are interested in  the effective string theory.

 Because of our ansatz, the metric is of the
Schwarzschild type,
\begin{equation}
          ds^2=-\left(1-\frac{2Gm}r\right){\rm e}^{-2\delta}dt^2
                      +\left(1-\frac{2Gm}r\right)^{-1}dr^2
                      +r^2 (d\theta^2 +\sin^2\theta d\phi^2).
\label{2-30}
\end{equation}
The mass function $m=m(r)$ and the lapse function $\delta=\delta
(r)$ depend on only the radial coordinate $r$.

Varying the action (\ref{2-10}) and substituting ans\"atze 
(\ref{3-80}) and (\ref{2-30}), we find the field equations;
\begin{eqnarray}
& & \delta ' = h^{-1} \left[ -\frac{1}{2} \tilde{r} \phi^{\prime 
2} 
            -\frac{{\rm e}^{-\gamma \phi}}{4\tilde{r}}
            \left\{ {\rm e}^{2\delta}              B^{-2}
             \tilde{a}^2 w^2+w^{\prime 2} 
             +\frac{2\tilde{m}}{\tilde{r}}
             (\gamma^2 \phi^{\prime 2}
             -\gamma \phi^{\prime \prime} )
          \right\} \right],
\label{3-100}  \\
& & \tilde{m}'= h^{-1} \left[ \frac{1}{4} 
           B
           \tilde{r}^2 \phi^{\prime 2} 
          +\frac{{\rm e}^{-\gamma \phi}}{16}
          \left\{ {\rm e}^{2\delta} \left(\tilde{r}^2 \tilde{a}^{\prime 
2}
          +2B^{-1} 
           \tilde{a}^2 w^2 \right)    
         +2Bw^{\prime 2}
          +\frac{(1-w^2)^2}{\tilde{r}^2}  \right. \right.  
\nonumber   \\
 & &  \;\;\;\;\;\;\;\; 
 \left. \left. 
    +4B \frac{2\tilde{m}}{\tilde{r}}
           (\gamma^2 \phi^{\prime 2}
           -\gamma \phi^{\prime \prime} )  
          +8\gamma \phi' \frac{\tilde{m}}{\tilde{r}^2}
          \left(B-\frac{\tilde{m}}{\tilde{r}} \right)
           \right\}\right],
\label{3-110}  \\
& & \left[ {\rm e}^{-\delta} \tilde{r}^2
          B \phi' \right]'
   -\frac{{\rm e}^{-\gamma \phi}}{8}
       \gamma {\rm e}^{-\delta} \tilde{r}^2
       \left[{\rm e}^{2\delta} \left\{ \tilde{a}^{\prime 2} 
       +\frac{2  \tilde{a}^2 w^2}{\tilde{r}^2} 
      B^{-1} \right\}
       -\left\{ \frac{2  w^{\prime 2} }{\tilde{r}^2}
       B
      +\frac{(1-w^2)^2}{\tilde{r}^4} \right\}  \right.
\nonumber   \\
 & &  \;\;\;\;\;\;\;\; 
     \left. +\frac{4}{\tilde{r}^2}
      \left\{2f^2 +B (\delta' f +f')      \right\}
     +\frac{4}{\tilde{r}^2} (\delta' f -f') 
    \right] = 0,
 \label{3-120}  \\
& & \left[ {\rm e}^{\delta} \tilde{r}^2{\rm e}^{-\gamma \phi}\tilde
{a}' \right]'
       -2{\rm e}^{\delta} {\rm e}^{-\gamma \phi} B^{-1}
       \tilde{a} w^2 = 0,
\label{3-130}  \\
& & \left[ {\rm e}^{-\delta}  B {\rm e}^{-\gamma \phi} w' \right]'
       +{\rm e}^{\delta} {\rm e}^{-\gamma \phi} B^{-1}
       \tilde{a}^{\prime 2} w 
       +{\rm e}^{-\delta} {\rm e}^{-\gamma \phi}
       \frac{w(1-w^2)}{\tilde{r}^{2}} = 0.
 \label{3-140}  
\end{eqnarray}
Here we have used the dimensionless variables;
$\tilde{r}=r/\sqrt{\alpha'}, \;\;
     \tilde{m}=Gm/\sqrt{\alpha'}$ and
     $\tilde{a}=a \sqrt{\alpha'}$. 
A prime in the field equations denotes the derivative with respect 
to $\tilde{r}$, and
\begin{eqnarray}
B & =& 1-\frac{2\tilde{m}}{\tilde{r}},\\
h & = &   1+\frac{{\rm e}^{-\gamma \phi}}{2\tilde{r}} \gamma \phi' 
            \left(B-\frac{\tilde{m}}{\tilde{r}} \right),  \\
f & = & h^{-1} \left[
\frac{\tilde{m}}{\tilde{r}^2} + \frac{1}{4}   B
           \tilde{r} \phi^{\prime 2} 
          -\frac{{\rm e}^{-\gamma \phi}}{16\tilde{r}}
          \left\{{\rm e}^{2\delta} \left(\tilde{r}^2 \tilde{a}^{\prime 2}
          -2B^{-1} \tilde{a}^2 w^2 \right) -2Bw^{\prime 2}
          +\frac{(1-w^2)^2}{\tilde{r}^2}  \right\} \right] .
\label{3-150} 
\end{eqnarray}
Note that
those equations can be applied for a $U(1)$ gauge field 
by setting
\begin{eqnarray}
   w & \equiv & \pm 1, \\
   \tilde{a}w & \equiv & 0 , \label{3-15}
\end{eqnarray}
with $\tilde{a}$ being a non-trivial potential.
The latter
condition (\ref{3-15}) corresponds to the vanishing of  the
self-interaction due to the non-Abelian  term.

As for boundary conditions for the metric functions on the event
horizon and at spatial  infinity, we impose following three 
ans\"atze: 

(i) Asymptotic flatness at spatial infinity\cite{chuu1}, 
i.e., as $r \to \infty$,
\begin{eqnarray}
m(r) & \to & M= {\rm finite}, \label{2-130} \\
\delta(r) & \to & 0  \label{2-140}.
\end{eqnarray}

(ii) The existence of a regular horizon $r_H$, i.e., 
\begin{eqnarray}
2Gm_H & = & r_H, 
\label{2-150} 
\\
\delta_H & < & \infty.
\label{2-160}
\end{eqnarray}

(iii) The nonexistence of singularities outside the event horizon, 
i.e., for $r>r_H$,
\begin{equation}
2Gm(r)  <  r. 
\label{2-170}
\end{equation}
The subscript $H$ is used for the values at the event horizon
$r=r_H$.
As for the field functions we have
\begin{eqnarray}
\phi & \to & 0, \label{2-111}  \\
a & \to & 0, \\
w & \to & \left\{
        \begin{array}{rl}
           \pm 1, & (\mbox{globally magnetically neutral solution}).  
\\
           0,        & (\mbox{globally magnetically charged solution}).
        \end{array}
           \right.   \label{3-105}
\end{eqnarray}
as $r \to \infty$ and impose their finiteness on the horizon.
These conditions guarantee that the total energy of the present system 
is finite\cite{chuu2}.
The boundary conditions of the field functions on the event horizon 
depend on the gauge group. Hence we will discuss them individually
in the next section.

%
%
\section{New Dilatonic  Black Holes with Gauss-Bonnet Term}
\label{sec:Dilatonic_Black Hole}

In this section we present the solutions of the field equations 
(\ref{3-100})$\sim$(\ref{3-140}). We classified them into four 
types by their gauge charge, i.e. neutral, electrically
charged, magnetically  charged and ``colored'' black holes.
All  solutions need numerical analysis.

\subsection{Neutral Black Hole}

First we consider the case without any gauge field, which is
the simplest. The solution is a
Schwarzschild type black hole modified by the GB
term coupled to a dilaton field.  Then the  gauge field should
be set to
\begin{eqnarray}
   \tilde{a} & \equiv & 0,  \\
   w & \equiv & \pm 1,
\end{eqnarray}
which satisfies the field equations (\ref{3-130}) and (\ref{3-140}). 
From Eq. (\ref{3-120}),  we find the following relation for the
dilaton field on the event horizon
\begin{equation}
\phi^{\prime 2}_H - \frac{\phi'_H}{A\gamma} +3 =0 ,
\label{2-240} 
\end{equation}
where $A=e^{-\gamma \phi_H}/4\tilde{r}_H^2$. The 
quadratic equation (\ref{2-240}) has two roots as 
\begin{equation}
\phi'_{\pm} = 
        \frac{1\pm \sqrt{1-12A^2\gamma^2 }}{2A\gamma}.
\label{2-250} 
\end{equation}
Hence, for each value of $\phi_H$ at a fixed
 event  horizon $\tilde{r}_H$, we have two possible boundary values 
$\phi'_{\pm}$.

We integrate the field equations (\ref{3-100})$\sim$(\ref{3-120}) 
from the horizon $r=r_H$ with the boundary conditions 
(\ref{2-150})$\sim$(\ref{2-170}) and (\ref{2-250}). Since the
equation of  the dilaton field (\ref{3-120}) becomes singular on the
event  horizon, we expand the equations and variables by power
series of $\tilde{r}-\tilde{r}_H$ to guarantee the regularity at
the horizon, and use their analytic solutions  for the first step
of  integration. We show the behavior of the field functions of
neutral black holes  with three different radius of event horizon in
Fig. 1. We find black hole solutions with a regular horizon only
when  we choose $\phi' _H =
\phi'_+$.

For smaller black holes, the dilaton field  varies more
rapidly  than that of the larger ones. This means that 
stringy  effects become more important for a smaller black hole,
as we expected. 
We see that the mass function decreases first near 
the horizon and then increases afterward, approaching a finite value.
There is  a region where the effective energy density, which is
defined by $(dm/dr)/4\pi r^2$, becomes
negative.  From Eq. (\ref{3-110}), 
$\tilde{m}'$ is evaluated on the event horizon as
\begin{equation}
\tilde{m}'_H= -\frac{A\gamma\phi'_H}{2(1-A\gamma\phi'_H)} 
 =  -\frac{1}{6} 
     \phi^{\prime 2}_H ,
\label{2-260} 
\end{equation}
which is  negative definite. 
Hence the  function $m$ always decreases in the vicinity of the event
horizon.  It is one of the essential points for existence of the
neutral black  holes that the energy density becomes
negative\cite{KMRTW}. Regarding the GB term
as a source term of the Einstein equations and applying the similar
analysis that was used to prove the no-hair theorem for 
a scalar field, we  can show
that if the time-time component of the energy-momentum tensor (the
effective energy density) is positive  everywhere, no non-trivial
solution can exist. However in our situation this is not  the case.
 So
we find  a new solution even if the gauge field  is absent.

We show $M-r_H$ relations  in Fig. 2. Note that  there is an end
point for each branch, where $\phi'_+$ and $\phi'_-$  coincide. 
With  this fact, we can  prove that
$\phi^{\prime \prime}_H$ and 
$\delta^{\prime}_H$ diverge.  We have also shown 
that
$R_ {\mu\nu\rho\sigma} R^{\mu\nu\rho\sigma}$ diverges on the event 
horizon, which means a naked singularity appears at the end point
of the branch\cite{KMRTW}. We shall call  this point the singular point 
S.

Near the singular point, we also  find a critical point C, 
which gives a
lower bound for the black hole mass, i.e., below which mass no
solution  exists (see Fig. 3(b)).  Since this solution is a modification
of Schwarzschild black hole, we find no other black hole solutions
without a gauge field. 
No naked singularity appears for the solution at
the  critical point. We have two 
black hole solutions in the mass range of $M_C<M<M_S$, where 
 $M_C=6.02771M_{PL}/g$ and $M_S=6.02813M_{PL}/g$ are the masses 
of 
the
critical and singular solutions, respectively. As we will see later,
the stability of these black holes changes at the critical point C
indicates that the singular end point solution becomes
unstable. 
 When we discuss the
 evolution of the black holes, it will also
important be that there exists a regular critical solution whose mass
is smaller than that of the singular solution

\subsection{Electrically Charged Black Hole}

Electrically charged black hole solutions with the U(1) gauge field 
can be obtained by setting
$   w  =  \pm 1, 
   \tilde{a}w  =  0$
in Eqs. (\ref{3-100})$\sim$(\ref{3-140}).
Then the electric potential $ \tilde{a}^{\prime \prime}$  is
integrated once to give
\begin{equation}
     \tilde{a}' ={\rm e}^{-\delta} {\rm e}^{\gamma \phi} 
           \frac{Q_e}{\tilde{r}^2},
\label{2-120}
\end{equation}
where $Q_e$ is a constant of integration and denotes a
normalized  electric charge.  The physical charge is given as
$gQ_e$.

From Eq. (\ref{3-120}), we find the following relation for the
dilaton field on the event  horizon
\begin{eqnarray}
& &  2A^4\gamma^4 e^{2\gamma\phi_H}Q_e^2 \phi_H^
{\prime 3}
- \left[A^2 \gamma e^{2\gamma\phi_H} Q_e^2  
       \left(2A^2\gamma^2
      +5A\gamma^2+1\right)-2A\gamma
      \right] \phi_H^{\prime 2}
\nonumber \\
& & \;\;\;\; + \left[A^4\gamma^2 e^
{4\gamma\phi_H}Q_e^4
     + A e^{2\gamma\phi_H}Q_e^2 
     \left(2A^2 \gamma^2+4A\gamma^2
      +1 \right) -2 \right] \phi'_H
\nonumber \\
& & \;\;\;\; + \left[\frac{A^3 \gamma}{2} e^
{4\gamma\phi_H} Q_e^4
     -A\gamma e^{2\gamma\phi_H} Q_e^2 
     \left(6A+1\right) +6 A\gamma  \right] =0 .
\label{2-180}
\end{eqnarray}
There are three roots of $\phi'_H$, $\phi'_1<\phi'_2<
\phi'_3$. We can,
however,  obtain a regular solution only  for
$\phi'_H=\phi'_2$. 
This is understood from the fact that in the limit of $Q_e
\rightarrow 0$, we recover the previous condition (\ref{2-240}) and
$\phi'_3=\infty$.  We plot the field functions of  the solutions
with $Q_e=1.0$ in Fig. 3. Their behaviors are almost same as those
of the neutral case  qualitatively.

We show $M-r_H$ relations of black holes with $Q_e=0.4$ 
and $1.0$  in Fig. 2. As the neutral black hole case, we also
find the critical  and the singular points (the end point on the
branch), C and S.  Fig.
2(c) is a magnification around the critical  point  C. 
No singular behavior appears for the critical
solution and below the critical point, no
solution  exists.   At the singular point,  a naked singularity
appears.

The existence of a  critical mass is also known in a
Reissner-Nordstr\"om black hole  with a fixed charge. In that case
the outer and inner horizons  coincide and the black hole becomes
extreme at the critical mass.  Our solution curve with a constant
charge in $M-r_H$ diagram is also  vertical at the critical  point.
However, we will see in the next section  that our critical
point C has different thermodynamical properties from that of the 
extreme point in Reissner-Nordstr\"om black hole.

\subsection{Magnetrically Charged Black Hole}

Next we turn to a magnetically charged solution.
In the present  subsection we extend the gauge field from the U(1) 
to  SU(2) group. 
This can be done by setting
\begin{eqnarray}
   \tilde{a} & \equiv & 0,  \\
   w & \equiv & 0.
\end{eqnarray}
Note that it is not the 'tHooft-Polyakov type, which is  obtained
 through a spontaneous symmetry breaking by a Higgs field  as in
Refs. \cite{'tH,LNW,Bre,Aic,Tach}, but the Wu-Yang type 
solution\cite{WY}. 
It is a kind of the dual solution of  U(1) electrically charged
one. The value of the magnetic  charge is quantized as
$Q_m=1.0$.

On the event horizon, Eq. (\ref{3-120}) becomes 
\begin{eqnarray}
&& \left[A^3 \gamma^3 \left(2A+1\right)
      +A \gamma \left(A-2\right) \right]\phi_H^{\prime 2} - \left[
A^2\gamma^2 \left(A^2+2A+ 2\right) 
          +  \left(A-2\right)\right] \phi_H' \nonumber \\
&& ~~~~~~~~~~~~~~~~~~~~~~ - \frac{A\gamma}{2} (A^2-12A+10)=0.
\label{3-115} 
\end{eqnarray} 
We have two roots of $\phi'_{\pm}$ but find again  a regular solution
only for $\phi'_H=\phi'_+$. The behavior
of the field functions is similar to that of neutral and
electrically charged solutions. We  plot the $M-r_H$ relation in
Fig. 4. We discuss its properties in the next
subsection together with the ``colored''  case.

\subsection{Dilatonic Colored Black Hole with Gauss-Bonnet Term}

Here we set the 'tHooft-Polyakov ansatz $\tilde{a}\equiv 0$, i.e., 
purely magnetic YM field strength exists. 
In the SU(2) Einstein-Yang-Mills system,  
the no-hair theorem for a
spherical monopole or dyons 
was proved\cite{NohairEYM1,NohairEYM2}. It states that there exists
no  static, spherically symmetric regular non-Abelian black
hole solution which has  non-zero global Yang-Mills magnetic
charge with or without an electric charge. Although it is not clear
whether  this kind of no-hair theorem holds in the present case, we
set
$a=0$  in this paper.

We obtain the following relation on the horizon from Eq. 
(\ref{3-120}),
\begin{equation}
 C_1 \phi_H^{\prime 2} + C_2 \phi_H' +C_3 =0 ,
\label{3-116} 
\end{equation} 
where
\begin{eqnarray}
C_1 & = & A^2 \gamma (1-w_H^2)^2 \left(2A^2\gamma^2
      +A\gamma^2+1\right)-2A\gamma,
\label{3-125} 
\\
C_2 & = & -A^4 \gamma^2 (1-w_H^2)^4
     - A (1-w_H^2)^2 
     \left(2A^2\gamma^2+2A\gamma^2
      +1\right) + 1,
\label{3-135} 
\\
C_3 & = & -\frac{A^3 \gamma}{2} (1-w_H^2)^4
     + A\gamma (1-w_H^2)^2
     \left(6A+1\right) -6 A\gamma .
\label{3-145} 
\end{eqnarray}
This has again two roots $ \phi'_{\pm}$,
 and we find a  non-trivial solution only with the value
$\phi'_ {+}$, as the same as before. The YM equation (\ref{3-140})
on the event horizon becomes
\begin{equation}
 w'_H = -\frac{w_H(1-w_H^2)}{2f(r_H)}.
\label{3-160} 
\end{equation} 
Then  $w'_H$ is also determined by
$\phi_H$  and $w_H$. Now we have only one shooting
parameter, $w_H$, which will be fixed by an iterative
integration with the boundary condition 
(\ref{3-105}). Here we   assume 
$w_H >0$ without loss of generality, since the field equations are 
symmetric for the sign change  of $w$.

With the above conditions, we solve Eqs. (\ref{3-100}) 
$\sim$ (\ref{3-120}) and (\ref{3-140}) numerically and find  discrete 
families of 
regular  black hole solutions which are characterized by the node 
number $n$ of the YM potential, just as for the colored black 
hole\cite{Vol}. We show some of the solutions with $n=1$ in Fig. 5. 
We shall call those solutions ``colored'' black holes as well. Since
the YM  field damps faster than $\sim 1/r^2$,  those black holes have
no  global color charge related to the gauge field, just like
the colored black hole. Here we use double-quotation marks to
distinct our new  solutions from the original colored black hole.
 The dilaton field and
metric functions are similar to those of other  solutions discussed
in the previous subsections. There is a region where the
effective energy  density becomes negative.

The characteristic feature is that the YM potential $w$ is almost 
scale invariant. This is recognized as follows. 
We shall normalize the variables by $\tilde{r}_H$ to see the scale
invariance, i.e.,
$\hat{r}=\tilde{r}/\tilde{r}_H$, $\hat{m}=\tilde{m}/\tilde{r}_H$,
$\hat{a}=\tilde{r}_H \tilde{a}$.
In the limit of $\tilde{r_H} \to \infty$ 
the field equations become
\begin{eqnarray}
& &     \delta^{\hat{\prime}} = -\frac{1}{2} \hat{r} \phi^{{\hat{\prime}} 
2}
     +O \left( \frac1{\tilde{r}_H^2} \right),
\label{3-170}  
\\
& &     \hat{m}^{\hat{\prime}} = \frac{1}{4} 
     \left(1-\frac{2\hat{m}}{\hat{r}} \right) \hat{r}^2 \phi^
{{\hat{\prime}} 2} 
     +O \left(\frac1{\tilde{r}_H^2}\right),
\label{3-180}  
\\
& &     \left[ e^{-\delta} \hat{r}^2 
     \left(1-\frac{2\hat{m}}{\hat{r}}\right) 
    \phi^{\hat{\prime}} \right]^{\hat{\prime}}
     +O \left(\frac1{\tilde{r}_H^2}\right) =0,
\label{3-190}  
\\
& &     \left[ e^{-\delta} e^{-\gamma \phi} 
     \left(1-\frac{2\hat{m}}{\hat{r}} \right) w^{\hat{\prime}}
   \right]^{\hat{\prime}}
     +e^{-\delta} e^{-\gamma \phi} \frac{w(1-w^2)}{\hat{r}^2}
     +O \left(\frac1{\tilde{r}_H^2}\right) =0.
\label{3-200}  
\end{eqnarray} 
A prime with a hat denotes the derivative with respect to $\hat{r}$.
Eqs. (\ref{3-170}) $\sim$ (\ref{3-190}) are decoupled from the YM
field and are the same as those for the Einstein gravity with a
massless  scalar field.  Because of the no-hair theorem, we find
just a Schwarzschild solution with $\phi=0$.   The YM field 
equation (\ref{3-200}) is then the same as that in a fixed
Schwarzschild background  spacetime, i.e., 
\begin{equation} 
   \hat{r} (\hat{r} -2 \hat{M}) w^{\hat{\prime} \hat{\prime}}
   +2\hat{M} w^{\hat{\prime}}
   +w (1-w^2) = 0.
\label{3-210}  
\end{equation}
The non-Abelian YM field can have a nontrivial configuration 
although it makes no contribution to the black hole structure.
Then the solution $w = w^{\ast} (\hat{r}) 
= w^{\ast} \left(\tilde{r}/\tilde{r_H} \right) = 
 w^{\ast} \left(r/r_H \right)$ is scale invariant. From our analysis
 this
scale  invariance is still found  approximately
even for small black holes. The  configuration of 
the YM potentials appears
to be described by almost the same 
function of
$r/r_H$.  Although the metric function $\delta$  for a small black hole
(e.g.  $\tilde {r}_H= 1.4007$) 
varies so rapidly that the above argument may
seem to be invalid. This may be understood as follows. 
$\delta$ is decoupled from other field functions even in the 
original basic equations. As a result the YM potential $w$ is 
not affected by $\delta$, (see Eq. (\ref{3-210})).
Furthermore if $m(r)$ and $e^{-\gamma \phi}$ are almost constant,
which is confirmed from our numerical solutions, then the equation
for $w(r)$ turns out to be Eq. (\ref{3-210}), resulting in a scale 
invariance of $w(r)$.
 A similar behavior is found for large black holes in the EYMD 
system\cite{Tor}.

We show the $M-r_H$ relations of the neutral, the magnetically 
charged ($Q_m =1.0$) and the ``colored'' black holes with node 
number $n=1$ in Fig. 6.  There exists a lower bound for its
mass in  each  branch.  Since the solution 
does not exist for $r_H \to 0$, our result 
 is consistent with \cite{DG}, which says that 
there is no particle-like solutions in the present system. 

Fig. 6(b) is a magnification around the critical point for the
magnetically charged and ``colored" black hole. Although the  neutral
and the electrically charged branches  have a turning point, which
is a critical point, 
 the magnetically charged and 
``colored'' branches do not. 
The lowest mass corresponds to the end point, where $\phi'_{\pm}$
coincide.

Although we have not proved it analytically, we expect that
a naked singularity
appears at this end point as  neutral solutions,  because  $\delta
'$ for the black hole
 near critical  mass  appears to  diverge on the
horizon  from  Fig. 4(d), and in our numerical solutions.
Hence the critical  point coincides with the singular point S.

%
%
\section{Thermodynamical property and Stability}
\label{sec:Thermodynamics}

In this section we investigate the thermodynamical properties of 
the dilatonic black holes with the GB term and then analyze their
stability.  We also discuss their fates via the Hawking evaporation
process.

\subsection{Temperature, Entropy and Stability}

The GM-GHS solutions and the non-Abelian black holes have an
interesting thermodynamical property. That is, a discontinuity 
of the heat capacity of the GM-GHS solution appears, depending on 
the coupling constant of the dilaton field $\gamma$. Its critical 
value is just what we use in this paper\cite{GM}. Similar 
properties were obtained in the EYMD system\cite{Tor}. Our new black
holes may possess similar interesting properties, which is the
reason why we investigate the thermodynamical properties here.

Although black hole 
thermodynamics in non-Einstein theories is not well 
studied, we can define the temperature and the entropy of the 
black holes in our model and show that they obey the first low of 
thermodynamics\cite{Wald}.
The Hawking temperature is given as 
\begin{equation}
    T = \frac{1}{4\pi r_H} e^{-\delta_H} \left[1-2 \tilde{m}'_H 
\right],
\label{4-10} 
\end{equation}
for the metric (\ref{2-30}). The inverse temperature $\beta \equiv 
1/T$ versus the gravitational mass is shown in Fig. 7. We show the 
branches of the neutral, the electrically charged ($Q_e=1.0$), the 
magnetically charged ($Q_m=1.0$) and the ``colored'' black holes 
with $n=1$. For comparison we also plot the GM-GHS solution, 
which have same temperature as Schwarzschild black holes in
$\gamma=\sqrt{2}$ case. Comparing the branch of the neutral black 
holes with that of Schwarzschild black holes, we can guess that  the
GB term has the tendency to raise the temperature. This can be 
confirmed by seeing other branches. For example the dilatonic 
colored black holes in the EYMD system have a lower temperature than 
that of the Schwarzschild black holes in the mass range of Fig. 7
\cite{Tor}. However the new solutions with ``color charge"  are under 
the 
branch of the Schwarzschild black holes. These behavior comes  from
the contribution of $\tilde{m}^{\prime}$ term of Eq. (\ref{4-10}).
$\tilde{m}^{\prime}$ has the minimum value
 $\tilde{m}^{\prime}=-1/2$ at the singular point in the neutral case. As
for the  electrically charged black hole we can see the effect of the GB
term in from a different point of view. The thermodynamical properties
of GM-GHS solutions changes drastically as the coupling constant 
$\gamma$ shifts across $\gamma=\sqrt{2}$. For $\gamma<\sqrt{2}$  
the
temperature is always lower than Schwarzschild black hole and  it
vanishes in the extreme limit, while for $\gamma>\sqrt{2}$ the 
temperature is always higher than Schwarzschild black hole and it 
diverges in the extreme limit. For $\gamma=\sqrt{2}$, which is
also  the
applicable value in our case, the temperature coincides with the 
Schwarzschild case and it is finite even in the extreme limit. 
However the behavior of the temperature of the electrically  charged
solutions is similar to the GM-GHS solution with $\gamma>
\sqrt{2}$. The critical value of $\gamma$ in the system including 
the GB term must be different from $\gamma=\sqrt{2}$ if such a
value exists at all.

The effect of the GB term also appears when we discuss the heat 
capacity. It is always negative for all new black holes found here
 in spite of 
the existence of the gauge field. Hence  the GB term produces a 
large effect on the thermodynamical properties.

We also calculate the entropy of our new solutions. Although the 
entropy is expressed by the quarter of the area of an event horizon 
in the Einstein theory, this relation cannot be applied in the 
non-Einstein theory. Here we adopt the entropy proposed by 
Wald\cite{Wald}, which originates from the  Noether charge of the 
system. 
This entropy has several desirable properties. For example, it can 
be defined in a covariant way  in all diffeomorphism-invariant 
theories, which includes the present model, and it obeys the first low 
of black hole thermodynamics for an arbitrary perturbation of a 
stationary black hole. 

The explicit form of the entropy is derived by
\begin{equation} 
     S=-2\pi \int_{\Sigma} E_R^{\; \mu\nu\rho\sigma} 
         \epsilon_{\mu\nu}\epsilon_{\rho\sigma}, 
          \label{en1}
\end{equation} 
where $\Sigma$ is the event horizon 2-surface, $\epsilon_{\mu\nu}$ 
denotes the volume element binormal to $\Sigma$ and the integral 
is taken with respect to the induced volume element on $\Sigma$. 
$E_R^{\; \mu\nu\rho\sigma}$ is defined as
\begin{equation} 
     E_R^{\; \mu\nu\rho\sigma}=
       \frac{\partial {\cal L}}{\partial R_{\mu\nu\rho\sigma}},  
 \label{en2}
\end{equation} 
where ${\cal L}$ is a Lagrangian density. Substituting the present 
model (\ref{2-10}) into (\ref{en1}) and (\ref{en2}), we obtain a 
simple form
\begin{equation} 
     S=\frac{A_H}{4} \left(1+\frac{\alpha'}{2r_H^2} e^{-\gamma 
\phi_H}          \right),
      \label{en3}
\end{equation} 
where $A_H$ is the area of the event horizon. 
This expression is also true for $\gamma=0$ case, in which we find
only a trivial Schwarzschild solution.  Although the GB term is
totally divergent for  $\gamma=0$ and does not give any
contributions to the field equations, the black hole entropy
defined here has the additional constant $\pi \alpha '/2$ via a
surface term. 
Fig. 8(a) is the plots of 
$S/S_{\rm Sch}$ for new solutions.
The entropy of neutral black hole is always larger than
that of the Schwarzschild solution with $\gamma=0$, 
although the area of neutral black holes is smaller than that
of Schwarzschild solutions with a same mass, from Fig. 2.
Since $\phi_H$ is negative for neutral solutions in Fig. 1(a),
then the second term of Eq. 57 make the entropy much larger.
Similarly, charged black holes have larger entropy than
that of Reissner-Nordstr\"om solutions.
We depict the $M$-$S$ diagram in Fig. 8(c), which shows 
 the same property as in the 
Einstein theory that the entropy becomes smaller  when the system 
includes the gauge field.

For the neutral and electrically charged case, we find the critical
point C and have two different black hole solutions with the same
mass.  Since the entropy may be a good indicator for the
stability\cite{Tor2,Tor3,Tach}, we show the fine structure of 
$M$-$S^{\ast}$ diagram near
the critical point in Fig. 8(b) for  the neutral black hole.  
We subtract a linear function which passes through an
appropriate point A and the end point S from
the entropy $S(M)$ in order to show the structure near the critical
point clearly.
Hence only the relative values on the vertical axis are significant.
We
find a cusp  structure at the critical point C. The cusp reminds us
of the  catastrophe theory. Catastrophe theory is a mathematical
tool established by Thom  to explain a variety of changes of state
in nature, in particular a  discontinuous change of states which
occurs eventually in spite of  gradual changes of parameters of a
system\cite{Thom}. It is widely  applied in various research fields
and in particular  we showed that this theory  is 
applicable to the stability
analysis of various types of non-Abelian black
holes\cite{Tor3,Tach}. 

As in the previous non-Abelian black
hole case, we adopt the mass and the entropy of a black hole as
a control  parameter and a potential function, respectively. Then the
existence of the cusp  structure shows that the stability changes at
this point. Then we can conclude that the {\it upper} branch, AC, is
more stable than {\it  lower} branch, CS, since  the
entropy is the potential  function. As a result, the singular point S
becomes unstable. Although the stability of the  upper branch
AC is not confirmed from  Fig. 8(c)  because the  catastrophe
theoretical method gives us only the relative  stability, 
we expect that it is stable since no other solution branch exists.
It is also understood from the following argument. 
 We know the stability of the  black hole
solutions in both EMD  and EYMD systems. As in Ref.
\cite{DG}, if we have a numerical 
constant $\beta$ in front of GB term,
we may have another control parameter $\beta$.
If we solve a series of  black hole solutions from 
$\beta = 0$ to
$1$, we find that our present solutions on the AC branch must be stable.

The two curves AC and CS  do not intersect with a non-zero angle but 
become tangent to each other. Since a temperature is expressed by 
$T=dM/dS$ from the first low of 
thermodynamics, the temperature is continuous at 
this critical point. This is consistent with Fig. 7. 

Finally, it  must be noted, however,  that it has not been
  proved whether the entropy  defined by Eq. (\ref{en1})
satisfies the second low of the black hole  thermodynamics. In
particular in our case  there is a region where  the effective
energy density becomes negative.

\subsection{Fate of Dilatonic Black Holes}

We find that the temperature of the black hole is finite
for all mass range. Hence we  expect that our black holes
will not stop evaporating via Hawking radiation.
Then we expect the following fates of our black holes:
When the neutral or the electrically charged black hole reaches the
critical mass C, then the  system must shift to another phase which 
cannot be described as a  static spherically symmetric regular black 
hole.
For the  magnetically charged and the ``colored'' black hole,
the evaporation  proceeds until the spacetime  reaches the singular
point S and  a naked singularity will eventually appear on the
hypersurface where the event  horizon was located.

However we have to study the evaporation process
more carefully.  Recall that the temperature of the extreme black
hole with
$\gamma >\sqrt{2}$ in the  EMD system is infinite. Although the
naive expectation is that this leads to an infinite emission rate, 
Holzhey and
Wilczek  showed that the potential, through which created particles
travel  away to infinity, grows infinitely high in the extreme
limit, and then it is expected that the emission rate could be 
suppressed
to a finite  value\cite{HHW}. However, it turns out that our
numerical analysis shows this expectation is  incorrect\cite{Koga}.
In  our case, if the potential barrier becomes infinitely large at
the  critical point C or at the singular point S, the emission rate
might  be suppressed to zero though the temperature of the black
hole  remains finite. Then
evaporation  stops and  the black hole cannot reach the critical
or the singular  point.
Therefore we have to 
calculate the  potential barrier for some field in the background of
our new solutions.

Here we examine a neutral  massless scaler field $\Phi$. It obeys 
the Klein-Gordon equation;
\begin{equation}
\Phi_{,\mu}^{\; ;\mu} =0.
\label{4-1}
\end{equation}
The scalar field is expanded in harmonics.  We study one mode of
\begin{equation}
\Phi = \frac{\chi(r)}{r} Y_{lm}(\theta, \phi) e^{-i\omega t}.
\label{4-2}
\end{equation}
Eq. (\ref{4-1}) becomes separable, and then the radial 
equation can be written as 
\begin{equation}
\left[ \frac{d^2}{dr^{\ast 2}} + \omega^2 - V^2 (r) \right] \chi(r^
{\ast}) =0,
\label{4-3}
\end{equation}
where $r^{\ast}$ is the tortoise coordinate defined as
\begin{equation}
\frac{d}{dr^{\ast}} \equiv 
           \left(1-\frac{2m}{r} \right) e^{-\delta} \frac{d}{dr}
\label{4-4}
\end{equation}
and $V^2 (r)$ is the potential;
\begin{equation}
V^2 (r) = \frac{e^{-2\delta}}{r^2} \left(1-\frac{2m}{r} \right)
      \left[ l(l+1) +\frac{2m}{r} -2m' -r\delta' \left(1-\frac{2m}{r} 
\right)
      \right].
\label{4-5}
\end{equation}
We calculate the potential of the new black holes.
We plot
the ratio  of the black hole temperature to the maximum  of the
potential for the neutral  solution (Fig. 9). We also plot the
Schwarzschild black hole case for  comparison. All plots for the
neutral solutions have larger values  than that of Schwarzschild black
holes.  This result means that the  evaporation process of the neutral
solution may be faster than in the  Schwarzschild case. Of course, the
width of the potential is another  factor which determines the
evaporation rate. However it does not  differ much from that of the
Schwarzschild black hole (see Fig. 10). Hence we  conclude that the
black hole continues to lose its mass, even when the solution 
approaches the critical mass point or the singular solution
one.  We then may be faced with a naked singularity, or a transition to
a new black hole state (in the
neutral or the  electrically charged case).
Since we have no black hole solutions below the critical mass, the
new state might be a time dependent evaporating black hole or just
a naked singularity, which was hidden behind the event horizon. 
It is an open question.

%
%
\section{Concluding Remarks}
\label{sec:Remarks}

We have studied dilatonic black holes with a GB term and found 
four new types of solutions, i.e., the neutral, the electrically
charged,  the magnetically charged and the ``colored''  black holes.
The  structures of these black holes and the field configurations are
almost the  same. This may be because the GB term becomes dominant
in the basic equations. 
As a result, the following  previously-unknown properties are found
with the GB term: \\[.5em]
(1) The effective
energy  density becomes negative near the event horizon, which may
be responsible for the existence of a new type of black hole.\\
(2)  The
neutral or  the electrically charged black holes have the critical
point C, below  which mass no solution exists, and the singular point
S, where a naked  singularity appears. \\
(3) For the magnetically charged or the
``colored''  black holes there is no critical point but only the
singular point S where the black hole has a finite minimum mass. \\
(4) The entropy is calculated.  It takes a minimum value at the
critical point C for the neutral and electrically charged black holes.
We also find a cusp structure in the $M$-$S$ diagram, which means
that the stability changes at the critical point C.  Since we do not
find a cusp structure for magnetically charged and ``colored" black
holes, their stabilities do not change.\\
(5)  The black hole temperature is always finite even at the
critical or the singular point.  The heat
capacity is always  negative, like the Schwarzschild black hole, even
for the  electrically or magnetically charged case.  This is
because the mass of the extreme black hole without the GB term, which
 is given by
$GM_{extreme}/\sqrt{\alpha '}=g^2 Q_e$ for a fixed charge, 
is much smaller than the lower
bound for the present black hole mass.  Then the
 charge per unit mass for our black holes is rather small.\\
(6) From the finiteness of the
temperature and the study of  the
effective potential for a massless scalar field, we may conclude that 
the  evaporation process will not be stopped even at the critical
point or the singular point. Hence it seems  inevitable that
the black hole shifts to another state such as a dynamical
evaporation phase or that a  naked singularity will appear.

We should mention several further points. 
There may be a dependence on the metric frame. In this 
paper we have worked only in the Einstein frame. However some 
results may change if we go into the string frame since the 
conformal transformation includes a nontrivial dilaton field. 
Therefore we studied the system again in the string frame.
For example, we plot the $M-r_H$ relation of the neutral black 
holes in the string frame in Fig. 11. It is known that the 
gravitational mass 
and the inertial mass are not equivalent in the string frame. 
Here we use the former in 
Fig. 11. We can find  similar structures; the existence of a 
critical point and a singular point. However, the present critical
point is not the same as that in the Einstein frame.  Since the
stability should not change in a choice of the frames, we expect that
the entropy will not take a minimum value at this critical point and
no cusp structure will appear in the $M_{string}$-$S_{string}$
diagram. From the consistency for stability analysis, the entropy
should be minimized  at the critical point in the string frame.

Next is that the solution 
curve becomes vertical at the critical point in the $M-r_H$ 
diagram. Hence the rate of change of the size of a black hole becomes 
infinite there, i.e., if a black hole emits just one particle by the
evaporation  process (in fact it is possible even at the critical
point because the temperature and the effective potential are finite),
the size of the black hole will change drastically. This means  that the
quasi-static approximation is broken and the thermodynamical 
approach 
may also break down near the critical point. Thus we have to study this
process taking the back reaction into account. 

Finally,  our 
results are obtained by using the model 
(\ref{2-10}) which includes only the leading terms of the 
expansion parameter $\alpha'$. This expansion may not be valid for 
Planck scale black holes. The existence and behavior of our black
holes near the critical or the singular point will be modified. 
Hence there is a possibility that  the critical  and singular point are
removed by taking into account  the higher or all orders in $\alpha'$. 
It might turn out that a particle-like solution found by Donets and
Gal'tsov for $\beta<0.37$ exists in a string theory ($\beta=1$), which
could be important in discussion about the final state of a black hole
evaporation or the information loss problem.

\vspace{.7cm}

-- Acknowledgments --

We would like to thank J. Koga and T. Tachizawa for useful 
discussion and R. Easther for his critical reading of our paper.
This work was supported partially by the Grant-in-Aid for 
Scientific 
Research Fund of the Ministry of Education, Science and Culture  
(No. 06302021 and No. 06640412), 
by the Grant-in-Aid for JSPS Fellows (No. 074767),
and by the Waseda University Grant 
for Special Research Projects.


\newpage
\begin{flushleft}
{Figure Captions}
\end{flushleft}

\vskip 0.1cm
   \noindent
\parbox[t]{2cm}{ FIG. 1:\\~}\ \ 
\parbox[t]{14cm}
{We plot the behavior of (a) the dilaton field (b) the mass function 
and (c) the lapse function of the neutral black holes for three 
different radii of event horizon; $r_H / \sqrt{\alpha'} = 1.5704$ 
(solid line), $= 1.9784$ (dotted line), $= 3.0556$ (dashed line). We 
find the region where the effective energy density becomes 
negative. $\delta'$ seems to 
diverge on the event horizon as the black hole gets smaller.
}\\[1em]
\noindent 
\parbox[t]{2cm}{ FIG. 2:\\~}\ \ 
\parbox[t]{14cm}
{(a) The mass-horizon radius diagram for the neutral black holes 
($Q_e=0$) and the electrically charged black holes ($Q_e=0.4, \; 
1.0$). There is an end point, where a naked 
singularity appears, for each curve. 
We also plot the Schwarzschild black hole and the
GM-GHS solution for comparison.
(b) and (c) is a magnifications around the 
singular point S of neutral black 
holes and electrically charged black holes respectively. 
We find the critical point C, below which no solution exists. 
}\\[1em]
\noindent 
\parbox[t]{2cm}{ FIG. 3:\\~}\ \ 
\parbox[t]{14cm}
{We plot the behavior of (a) the dilaton field (b) the mass function 
and (c) the lapse function of the electrically charged black holes 
with $Q_e=1.0$ for three different radii of event horizon; $r_H / 
\sqrt{\alpha'} = 1.52$ (solid line), $= 1.70$ (dotted line), $= 2.00$ 
(dashed line). The qualitative behaviors of these functions are
 almost same as 
those of the neutral case.
}\\[1em]
\noindent 
\parbox[t]{2cm}{ FIG. 4:\\~}\ \ 
\parbox[t]{14cm}
{We plot the behavior of (a) the dilaton field (b) the mass function 
and (c) the lapse function of the magnetically charged black holes 
with $Q_m=1.0$ for three different radii of event horizon; $r_H / 
\sqrt{\alpha'} = 1.3864$ (solid line), $= 1.8929$ (dotted line), 
$= 3.0101$ 
(dashed line). The qualitative behavior of these functions is
 almost same as 
that of the neutral and electrically charged case.
}\\[1em]
\noindent 
\parbox[t]{2cm}{ FIG. 5:\\~}\ \ 
\parbox[t]{14cm}
{We plot the behavior of (a) the dilaton field  
(b) the mass function (c) the lapse function and 
(d) the YM potential of the ``colored'' 
black holes for three different radii of event horizon; $r_H/ \sqrt
{\alpha'} = 1.4007$ (solid line), $= 1.8964$ (dotted line), $= 
3.0117$ (dashed line). The behavior of the YM potential $w$ hardly 
depends on the size of the black hole. The Yang-Mills field damps 
faster than $\sim 1/r^2$ so  ``colored" black holes have no global 
YM charge.
}\\[1em]
\noindent 
\parbox[t]{2cm}{ FIG. 6:\\~}\ \ 
\parbox[t]{14cm}
{(a) The mass-horizon radius diagram for the neutral , the 
magnetically charged  ($Q_m=1.0$) and ``colored'' black holes ($n=
1$). We plot the Schwarzschild black holes and GM-GHS solutions for 
comparison. There exists a lower mass for each branch. (b) is a 
magnification around the end points of magnetically charged and 
 ``colored" black holes. There is no critical point but
only the singular point, where a naked singularity appears.
}\\[1em]
\noindent 
\parbox[t]{2cm}{ FIG. 7:\\~}\ \ 
\parbox[t]{14cm}
{The mass-temperature relations for the neutral, the electrically 
charged ($Q_e=1.0$), the magnetically charged ($Q_m=1.0$)  and 
the ``colored'' black hole ($n=1$). The temperature remains finite in 
all mass range for each branch.
The effect of the GB term also appears in the
heat capacity, which is always negative for each branch in spite
of the existence of the gauge field.
}\\[1em]
\noindent 
\parbox[t]{2cm}{ FIG. 8:\\~}\ \ 
\parbox[t]{14cm}
{(a) We plot the ratio of the entropy for the neutral, the 
electrically 
charged ($Q_e=1.0$), the magnetically charged ($Q_m=1.0$)  and 
the ``colored'' black hole ($n=1$) to that for the Schwarzschild
black hole. Here we define the entropy of the Schwarzschild black 
hole by Eq. (\ref{en3}), so it contains the contributions
of a surface term. We plot the entropy of Reissner-Nordstr\"om
black holes by dashed line for comparison. The neutral and
the charged solutions have larger entropy than Schwarzschild
and Reissner-Nordst\"om black holes respectively.
(b) is the entropy for these four types of solutions. The 
entropy becomes small when the 
system includes the gauge field. This behavior is same 
in Einstein theory. 
(c) is the magnification around the 
critical point of the neutral black holes. 
We change the vertical axis $S\to S^{\ast}$ in order to show the 
structure clearly by subtracting the linear function, which passes 
though
the points A and S, from the original entropy $S$. Hence the absolute
value have no meaning.
We can find a cusp structure, 
which means that the stability change occurs at the critical point 
C.
}\\[1em]
\parbox[t]{2cm}{ FIG. 9:\\~}\ \ 
\parbox[t]{14cm}
{The ratio of the temperature to the maximum of the potential, 
through which particles travel, of the neutral black holes. The ratio 
of new solutions is always larger than that of the
Schwarzschild black hole. Hence the evaporation proceeds until the 
black hole reaches to the critical point.
}\\[1em]
\noindent 
\parbox[t]{2cm}{ FIG. 10:\\~}\ \ 
\parbox[t]{14cm}
{We plot the configurations of the potential of neutral massless
 scalar field around the neutral black holes with three different
radii of the event horizon; $r_H / \sqrt{\alpha'} = 1.5704$, 
$= 1.9784$ and $= 3.0556$. We also plot those of the Schwarzschild 
solution
 case with the same size of black hole as each neutral solution. The
configurations are similar to those of Schwarzschild solutions for all
horizon sizes.
}\\[1em]
\noindent 
\parbox[t]{2cm}{ FIG. 11:\\~}\ \ 
\parbox[t]{14cm}
{The mass-horizon radius diagram for the neutral black holes 
in the string frame. 
There are also both critical and  singular 
points. This critical point, however, does not coincide with 
that in Einstein frame.
}\\[1em]


\begin{thebibliography}{99}
\bibitem{chuu1}
In our analysis, we set 
\begin{eqnarray}
\delta _H & = & 0,   \nonumber  \\
\delta (r) & \to & \delta_{\infty} = {\rm const.}
 \;\;\; {\rm as} \;\;\; r \to \infty,   \nonumber 
\end{eqnarray}
instead of the condition (\ref{2-140}) in order to simplify the
numerical calculations. This difference in $\delta$ is recovered by
rescaling the time  coordinate as
$t \rightarrow t e^{\delta_{\infty}/2}$.
\bibitem{chuu2}
Since the value of the dilaton field $\phi$ is not fixed up to
a constant because the constant difference can be absorbed in
the radial coordinate by  rescaling, we
can relax the condition (\ref{2-111}). For  example, we can assume that
the dilaton field approaches to some unknown constant value
$\phi_{\infty}$ as
$r\to
\infty$. Introducing 
$\bar{\phi}$ defined by 
$\bar{\phi} \equiv \phi -\phi_{\infty}$, and rescaling the variables 
as 
$\bar{r}  =  {\rm e}^{-\gamma \phi_{\infty}/2} \tilde{r}$, 
$\bar{m}  =  {\rm e}^{-\gamma \phi_{\infty}/2} \tilde{m}$ and
$\bar{a}  =  {\rm e}^{\gamma \phi_{\infty}/2} \tilde{a}$,
we find our boundary condition.
However, 
in our numerical analysis, because we have fixed a charge of black
hole in each branch, we have to  search for the value of $\phi _H$  by
use of an iteration method such that  $\phi_\infty$ vanishes.
Otherwise  $a$ must be rescaled if $\phi_\infty \neq 0$,
giving a different value of charge.
 Hence $\phi_H$ is not  a free parameter but a shooting parameter in 
the
electrically  charged black hole case.
\end{thebibliography}
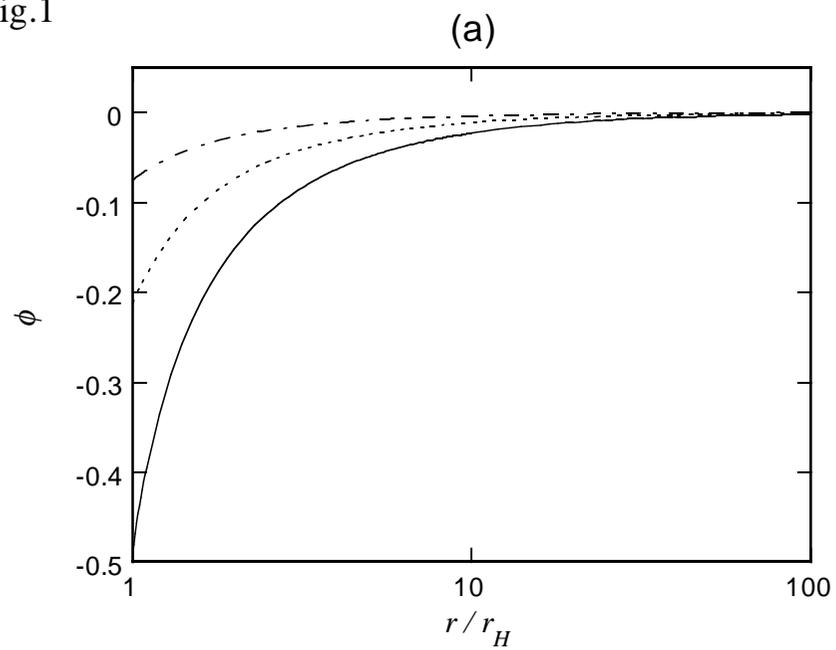
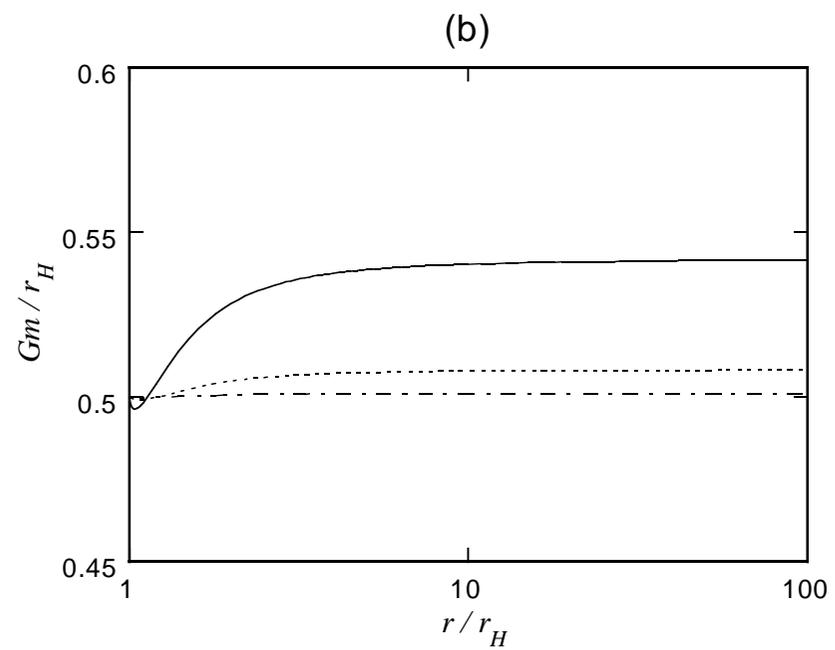
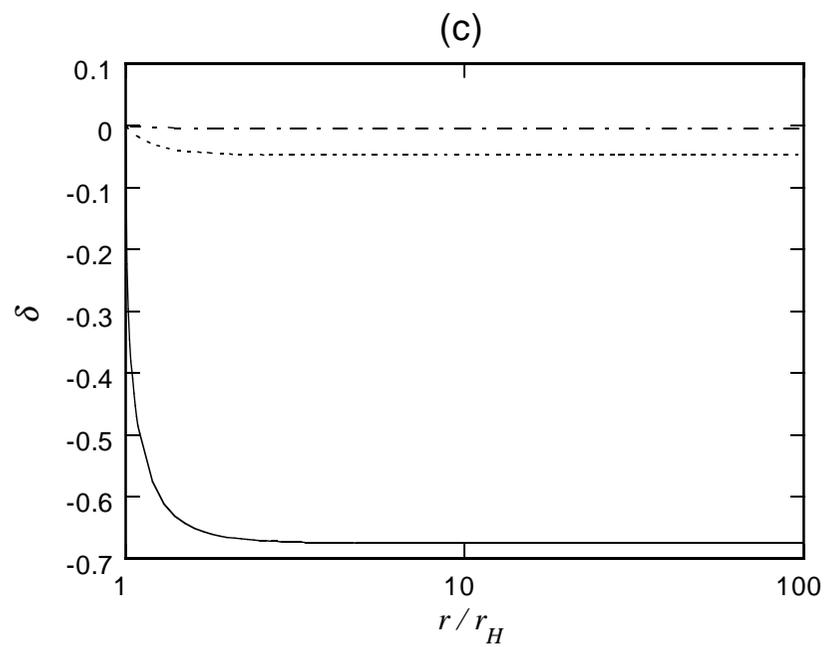
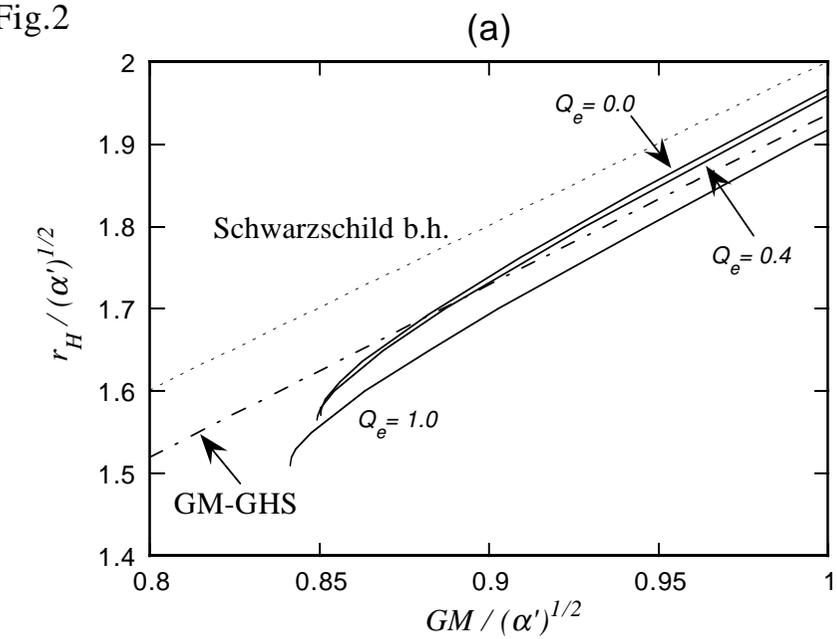
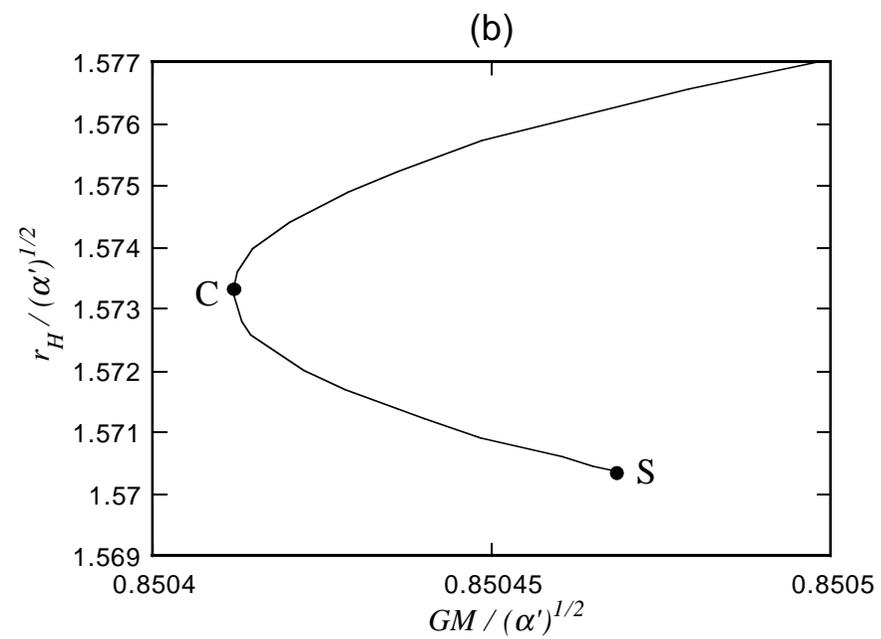
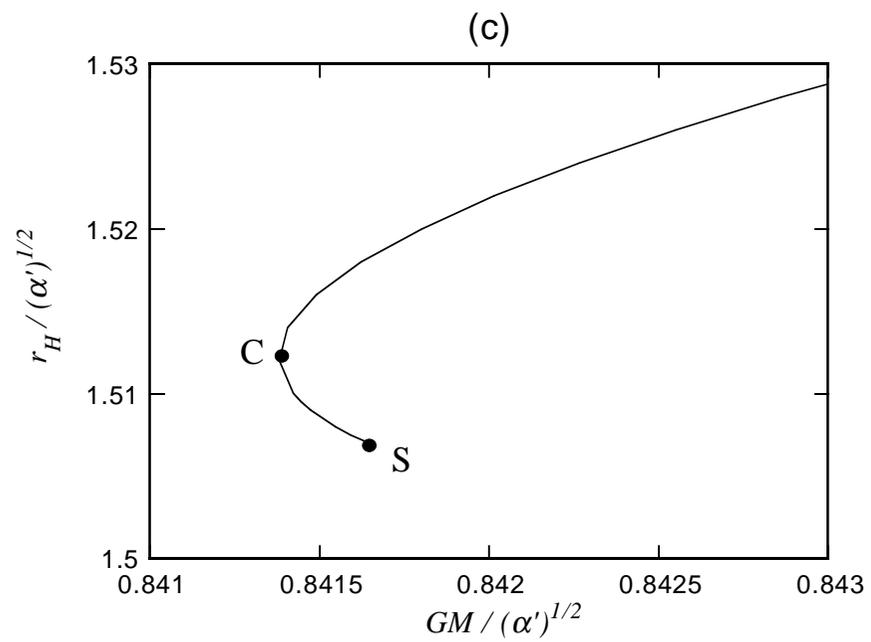
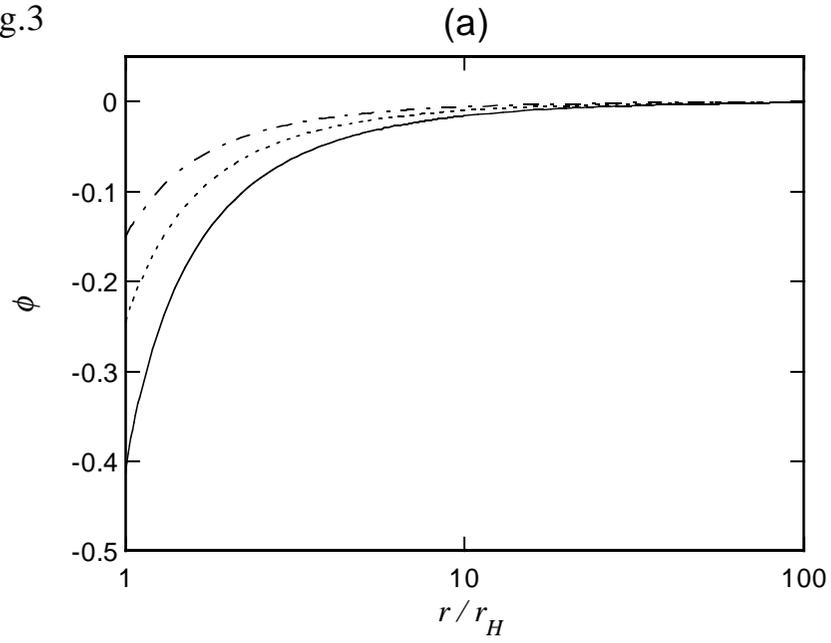
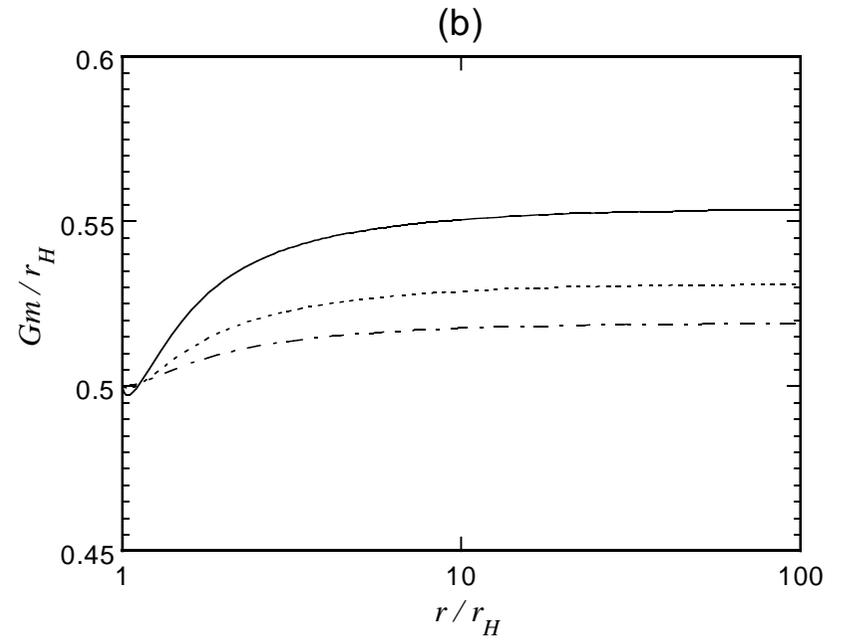
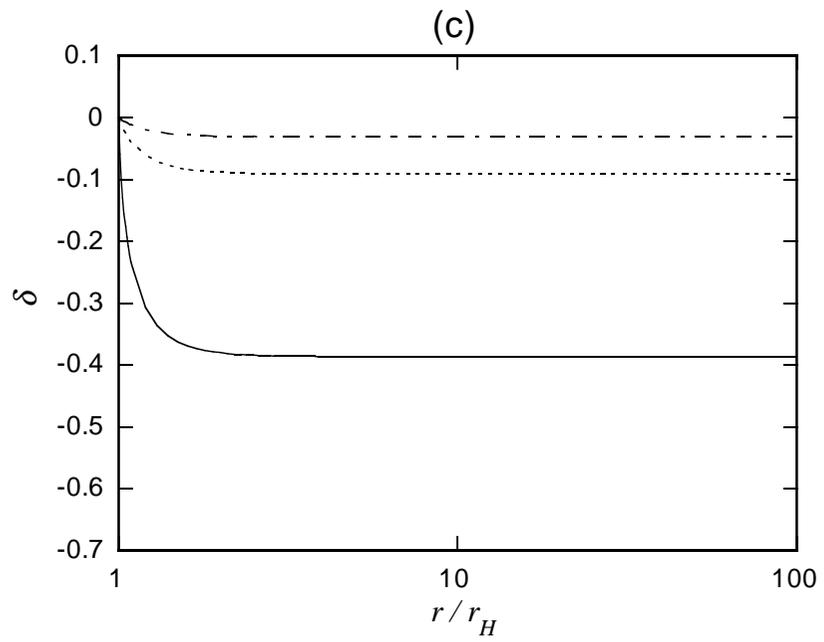
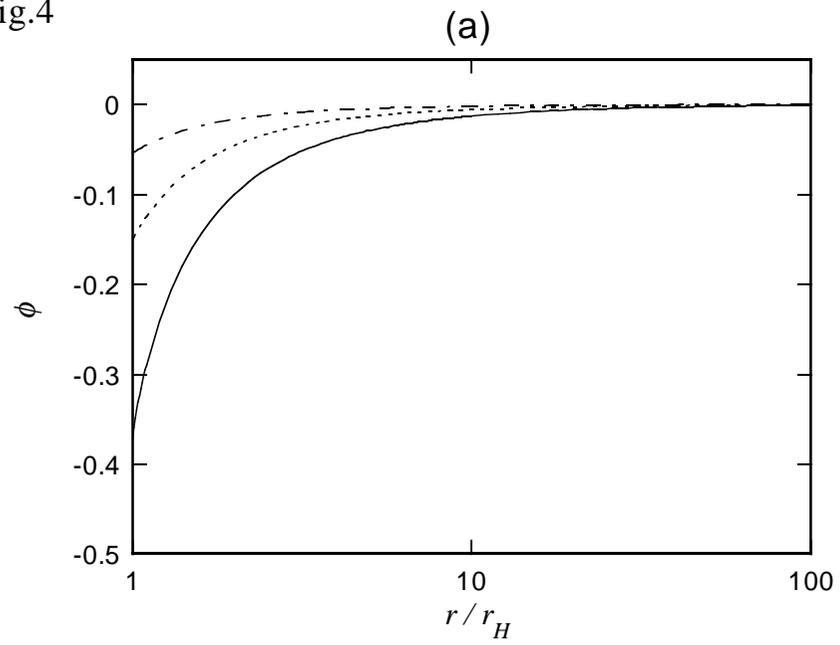
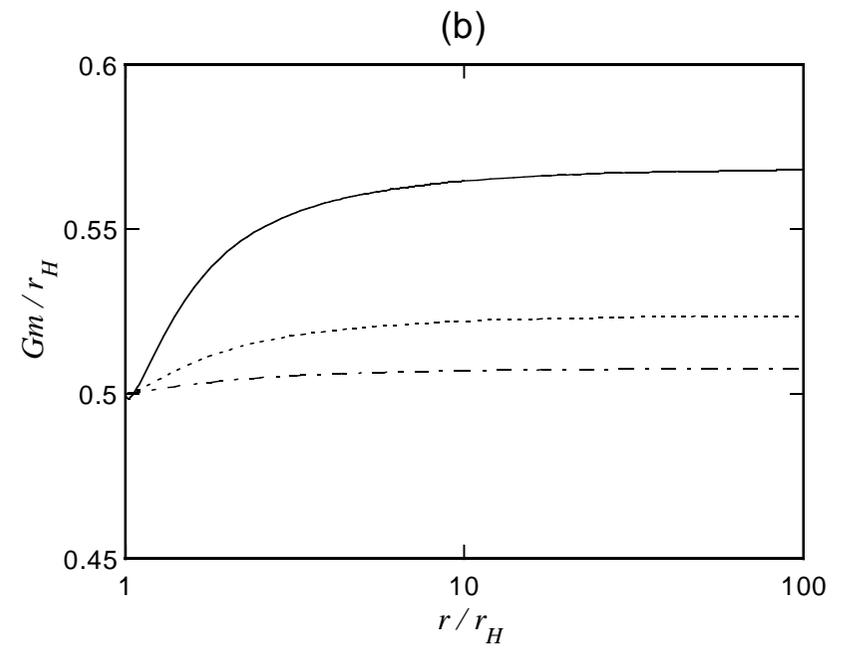
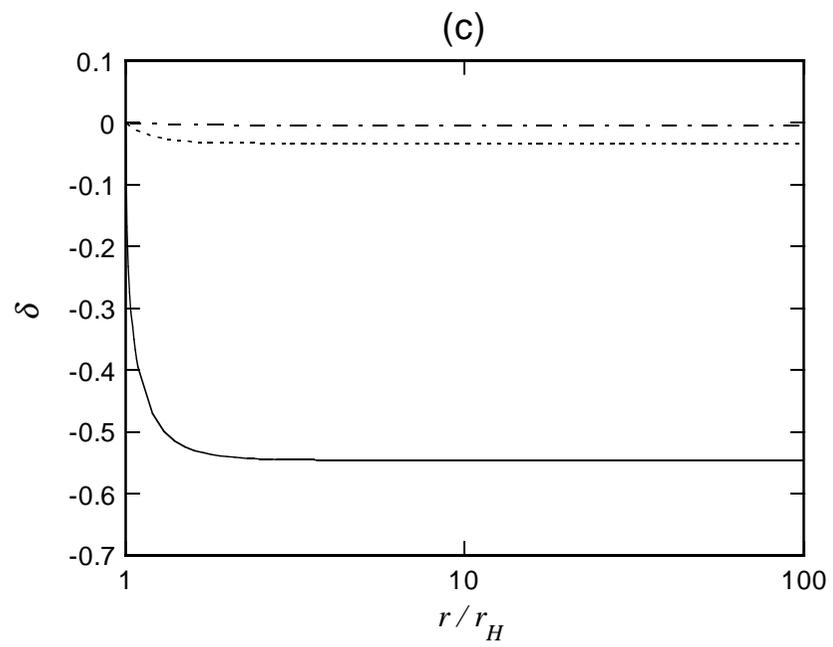
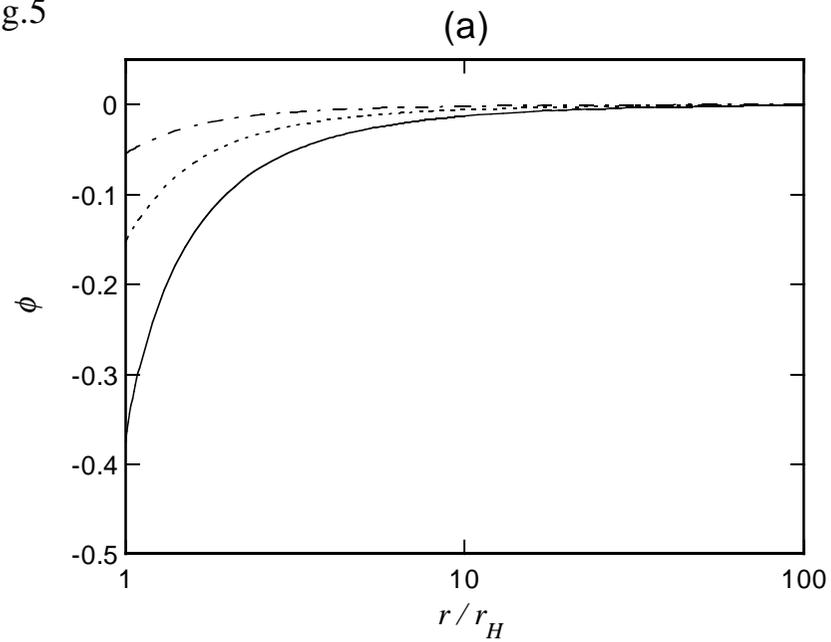
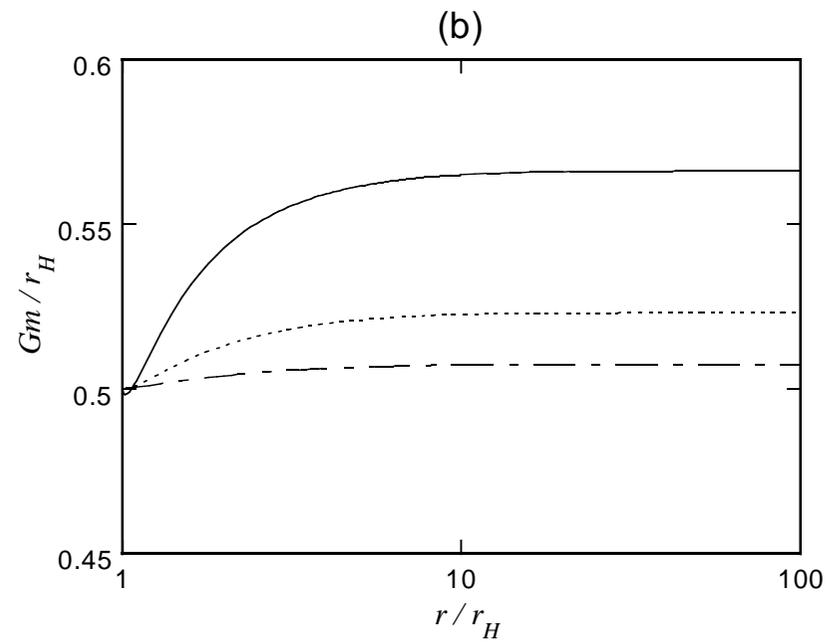
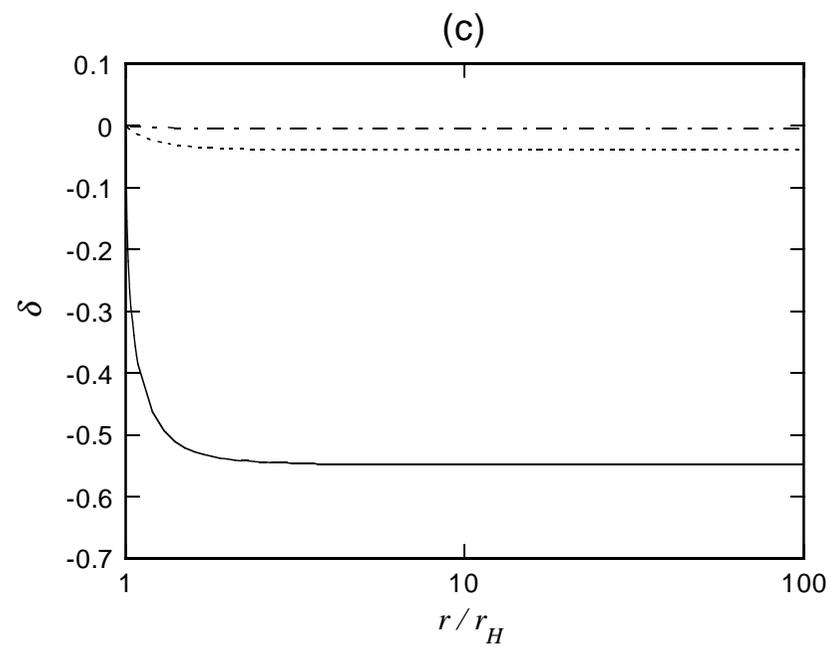
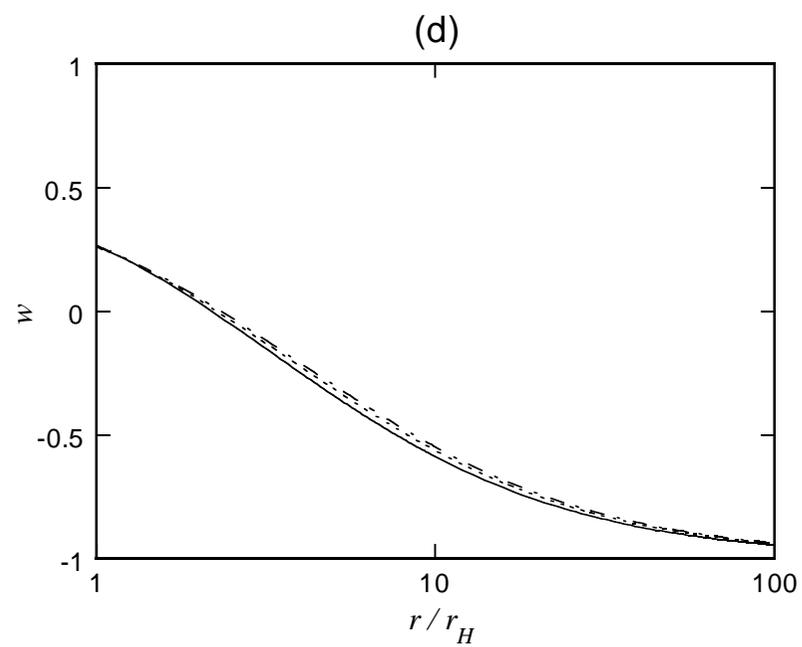
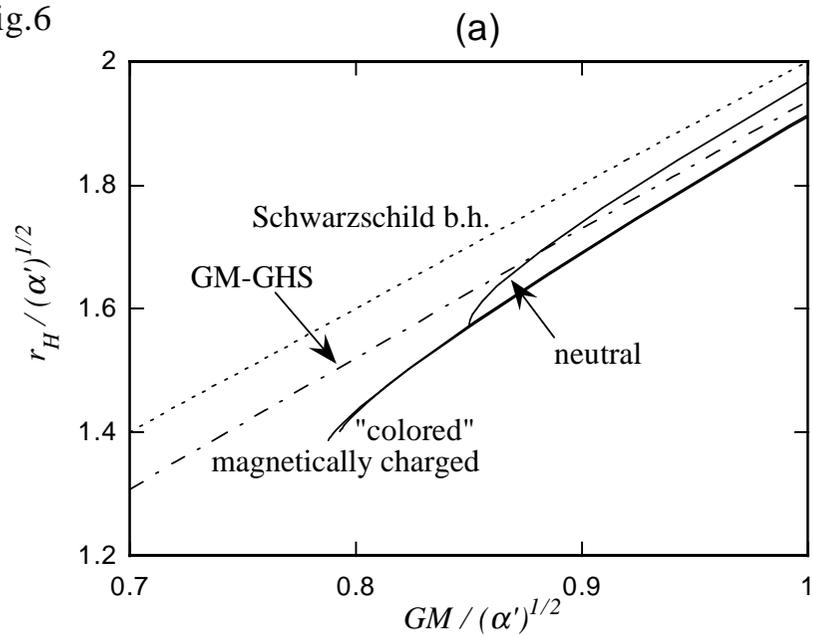
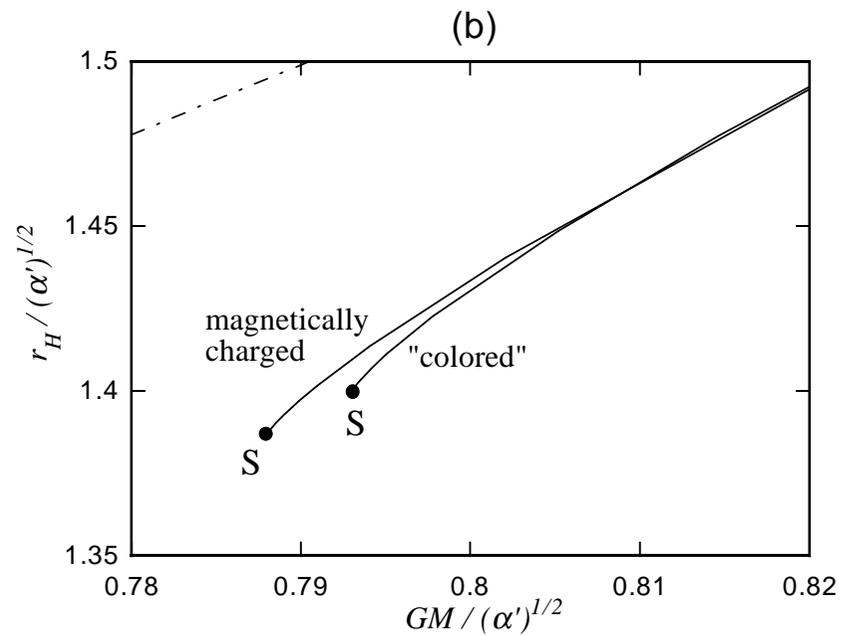
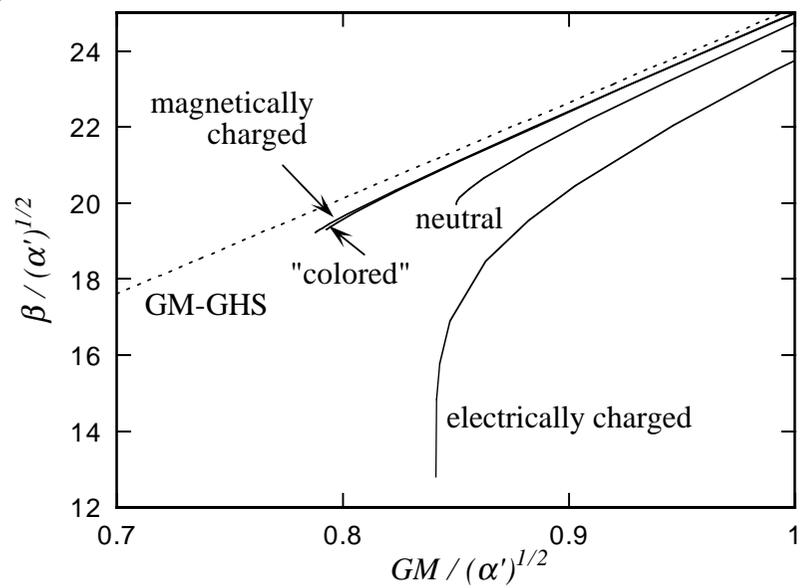
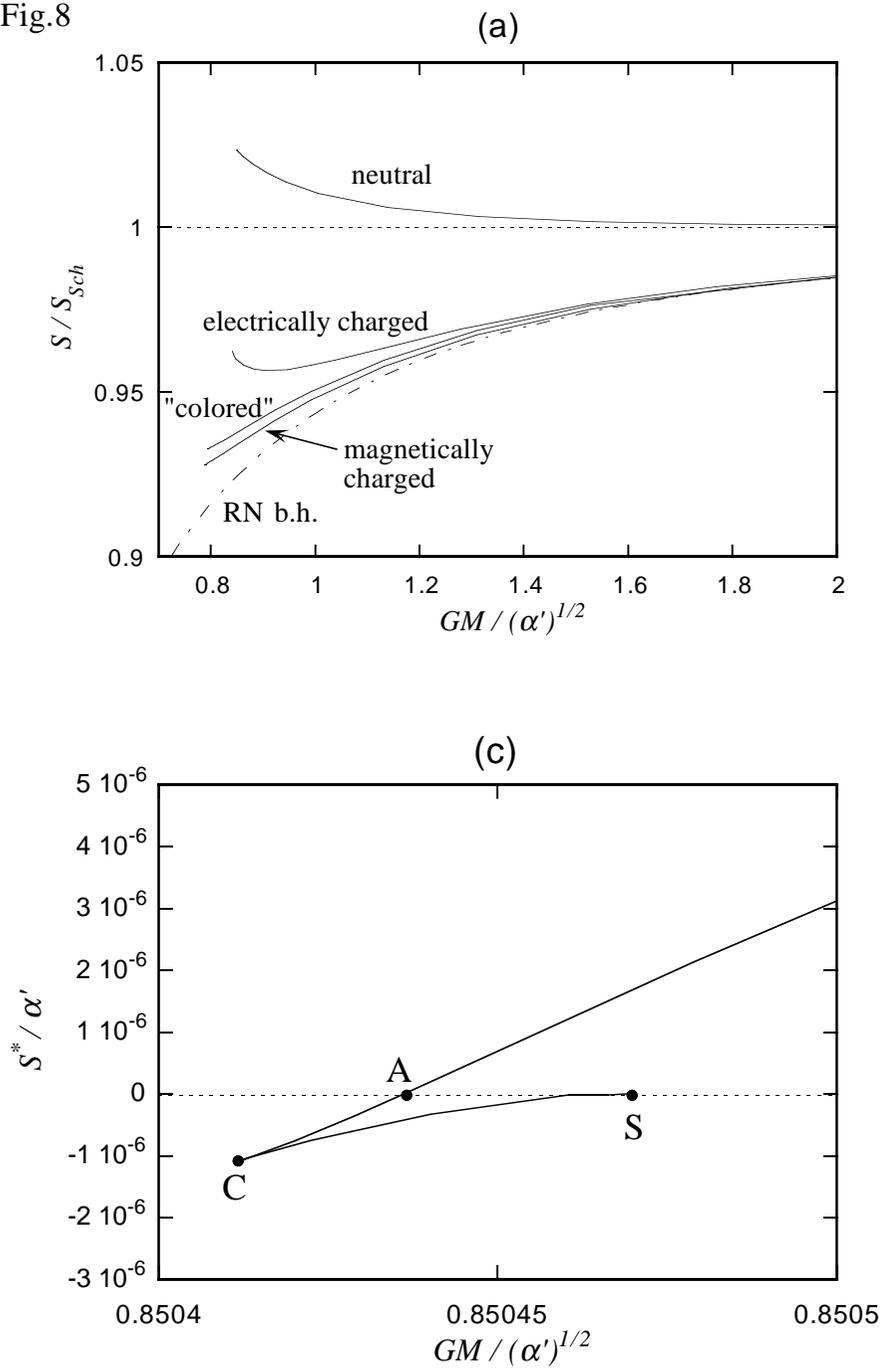
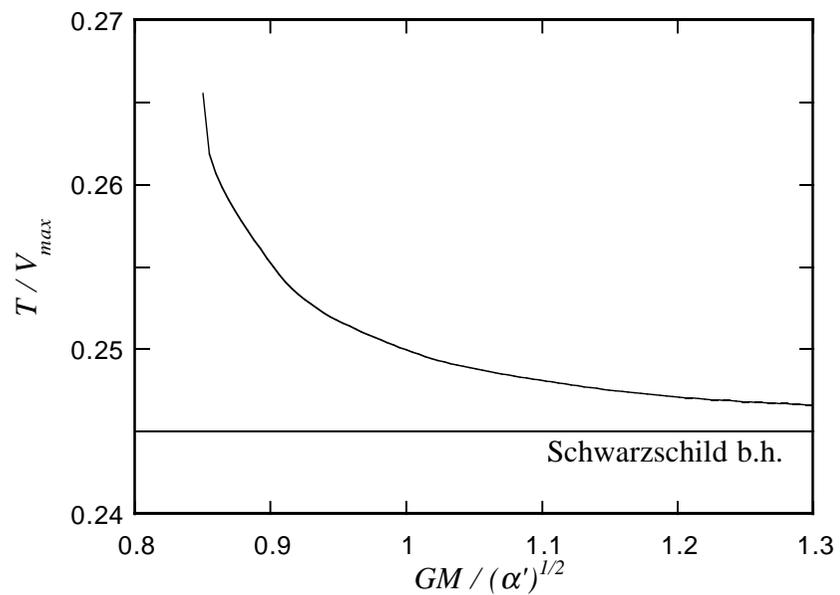
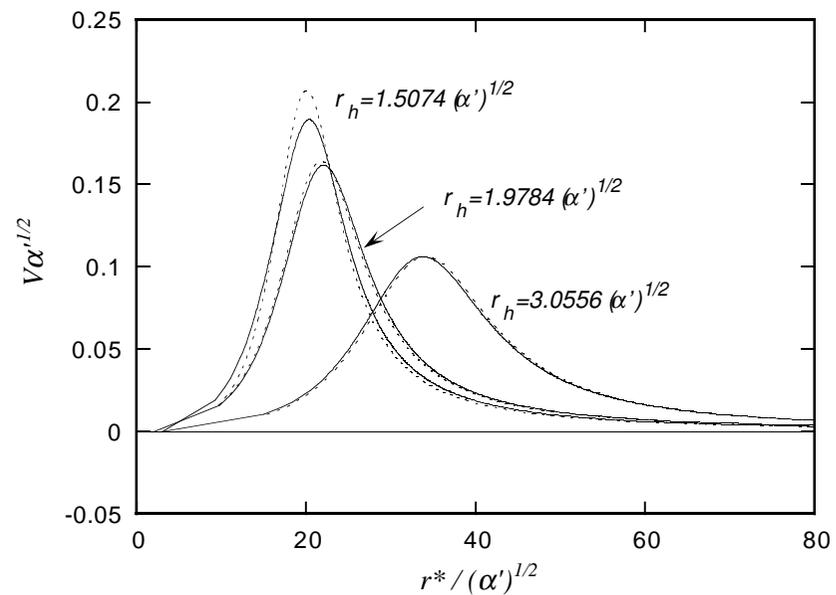
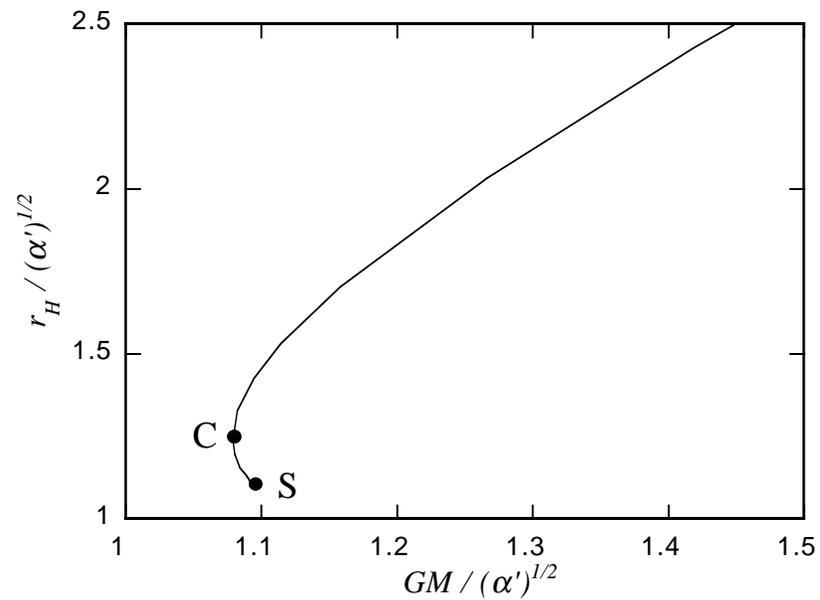
\end{document}